\newcommand{\be}{\begin{equation}}
\newcommand{\bea}{\begin{eqnarray}}
\newcommand{\ee}{\end{equation}}
\newcommand{\eea}{\end{eqnarray}}
\newcommand{\nn}{\nonumber}

\newcommand{\qa}{\alpha}
\newcommand{\qb}{\beta}
\newcommand{\qg}{\gamma}

\newcommand{\qe}{\varepsilon}

\newcommand{\qz}{\zeta}

\newcommand{\qy}{\theta}

\newcommand{\ql}{\lambda}
\newcommand{\qL}{\Lambda}

\newcommand{\qr}{\rho}
\newcommand{\qs}{\sigma}

\newcommand{\qf}{\varphi}
\newcommand{\qF}{\Phi}

\newcommand{\qj}{\psi}
\newcommand{\qJ}{\Psi}
\newcommand{\qo}{\omega}
\newcommand{\qO}{\Omega}

\newcommand{\tr}{{\rm tr}\,}

\newcommand{\tri}{\triangle}

\newcommand{\dagg}{^{\dag}}

\newcommand{\fr}[2]{{\textstyle \frac{#1}{#2}}}

\newcommand{\EE}{{\mathbb E}}

\newcommand{\one}{{\mathbb 1}}
\newcommand{\sH}{{\sf H}}
\newcommand{\bits}{ \{0,1\} }
\newcommand{\Hmin}{{\sf H}_{\rm min}}

\newcommand{\cB}{{\mathcal B}}

\newcommand{\cM}{{\mathcal M}}

\newcommand{\cQ}{{\mathcal Q}}
\newcommand{\cR}{{\mathcal R}}

\newcommand{\cX}{{\mathcal X}}
\newcommand{\cY}{{\mathcal Y}}

\newcommand{\vecv}{{\boldsymbol{v}}}

\newcommand{\vecn}{{\boldsymbol{n}}}

\newcommand{\vecE}{{\boldsymbol{E}}}

\newcommand{\vecX}{{\boldsymbol{X}}}

\newcommand{\vecsig}{{\boldsymbol{\qs}}}

\newcommand{\pr}{{\rm Pr}}

\newcommand{\isdef}{\stackrel{\rm def}{=}}

\newcommand{\ket}[1]{| #1 \rangle}
\newcommand{\bra}[1]{\langle #1 |}
\newcommand{\inprod}[2]{\langle #1 | #2 \rangle}

\newcommand{\pok}{p_{\scriptscriptstyle{\rm OK}}}

\documentclass[10pt]{article}
\usepackage{graphics}
\usepackage{graphicx}
\usepackage{url}
\usepackage{amsmath}
\usepackage{amssymb}
\usepackage{bbold}
\usepackage{a4wide}
\usepackage{enumitem}

\begin{document}

\newtheorem{theorem}{Theorem}[section]
\newtheorem{corollary}[theorem]{Corollary}
\newtheorem{property}[theorem]{Property}
\newtheorem{lemma}[theorem]{Lemma}
\newtheorem{definition}[theorem]{Definition}
\newtheorem{example}[theorem]{Example}
\newtheorem{conjecture}[theorem]{Conjecture}

\title{Optimal attacks on qubit-based Quantum Key Recycling}

\author{Daan Leermakers and Boris \v{S}kori\'{c} \\
{\tt\footnotesize d.leermakers.1@.tue.nl, b.skoric@tue.nl} }

\date{ }

\maketitle

\setlength{\parindent}{0mm}

\begin{abstract}
\noindent
Quantum Key Recycling (QKR) is a quantum-cryptographic primitive
that allows one to re-use keys in an unconditionally secure way.
By removing the need to repeatedly generate new keys
it improves communication efficiency.
\v{S}kori\'{c} and de Vries recently proposed a QKR scheme
based on 8-state encoding (four bases).
It does not require quantum computers for encryption/decryption
but only single-qubit operations. 
We provide a missing ingredient in the security analysis of
this scheme in the case of noisy channels: accurate bounds on the privacy amplification. 
We determine optimal attacks against the message and against the key,
for 8-state encoding as well as 4-state and 6-state conjugate coding.
We show that the Shannon entropy analysis for 8-state encoding reduces to 
the analysis of Quantum Key Distribution, whereas 4-state and 6-state suffer
from additional leaks that make them less effective.
We also provide results in terms of the min-entropy.
Overall, 8-state encoding yields the highest capacity.
\end{abstract}

\section{Introduction}
\label{sec:intro}

\subsection{Quantum Key Recycling}
\label{sec:introQKR}

Quantum communication differs significantly from classical communication.
On a classical channel it is trivial to read and copy all messages.
On a quantum channel, on the other hand, any form of eavesdropping is detectable.
This fact has been exploited by cryptographers since the 1980s,
most notably by the introduction of Quantum Key Distribution (QKD).
However, even before the invention of BB84 
another concept was studied: information-theoretically secure re-use of encryption keys.
If Bob detects no disturbance on the quantum channel, it may be safe to re-use 
the encryption key, in stark contrast to e.g.\;One Time Pad (OTP) encryption on a classical channel.
This idea was proposed in the paper
{\it ``Quantum Cryptography II: How to re-use a one-time pad safely even if P = NP''} 
\cite{BBB82} by Bennett, Brassard and Breidbart in 1982.
However, after the discovery of QKD the idea of Quantum Key Recycling (QKR)
received very little attention for several decades.
The thread was picked up again in 2003 by Gottesman \cite{uncl}
and
in 2005 by Damg{\aa}rd, Pedersen and Salvail \cite{DPS2005,DBPS2014}.
Gottesman's Unclonable Encryption offers a limited re-usability of key material.
Damg{\aa}rd et al
introduced a full key re-use scheme based on mutually unbiased bases in high-dimensional
Hilbert space.
A drawback of their scheme is that it requires a quantum computer to perform
encryption and decryption.
In 2016 Fehr and Salvail \cite{FehrSalvail2017}
and \v{S}kori\'{c} and de Vries \cite{SdV2016}
returned to qubit-based schemes that do not require a quantum computer.
Fehr and Salvail \cite{FehrSalvail2017} used BB84 states and introduced a new proof technique.
Their scheme is provably secure when there is very little channel noise.
\v{S}kori\'{c} and de Vries \cite{SdV2016} showed that it is advantageous to switch from 
4-state conjugate coding to 8-state encoding, and that 8-state encoding is equivalent to applying the
Quantum One Time Pad (QOTP) \cite{AMTW2000,Leung2002,BR2003}.
Their scheme is designed to work at similar noise levels as QKD.
The proof technique of \cite{FehrSalvail2017} can be directly applied to it,
but needs an accurate bound on the required amount of privacy amplification,
which was provided only for the noiseless case.

The long neglect of QKR is undeserved.
In a QKD-equipped world, QKR has an important role to play.
The process of repeatedly generating new QKD keys and then using them up
with classical OTP encryption is very wasteful of bandwidth.
One QKD instance followed by repeated
QKR runs is more communication-efficient.

\subsection{Contributions and outline}
\label{sec:contrib}

\begin{itemize}
\item
We determine optimal attacks against individual qubits in qubit-based QKR,
such that Eve introduces channel noise parametrised by the bit error rate $\qb$.
We apply the standard Shor-Preskill technique \cite{SP2000} to reformulate state preparation as
a measurement on an EPR state.
We apply noise symmetrisation \cite{RGK2005} to Alice and Bob's noisy EPR state,
followed by purification to obtain a worst-case description of Eve's ancilla state.
We find optimal POVM measurements by which Eve extracts from her ancilla 
information about the plaintext, as well as POVMs for attacking the key in
the known-plaintext setting.
We obtain POVMs for Shannon entropy as well as min-entropy.
\item
From the optimal POVMs we determine how much privacy amplification is needed:
this is dictated by the most powerful attack.
We find that it depends on $\qb$ which attack `wins'.
\begin{itemize}
  \item {\bf Shannon entropy}.
For 4-state and 6-state encoding, the winning attack at low $\qb$ is 
Eve stealing all qubits and performing a measurement to
estimate the plaintext.\footnote{
This is due to the fact that conjugate coding is not a particularly good encryption.
} 
At larger $\qb$, Eve collects ancillas from many QKR rounds and then performs a measurement 
on all the ancillas that are protected by the same basis key;
we show that this attack is (asymptotically) as powerful as the optimal qubit-wise attack
on QKD \cite{Bruss1998}.
For 8-state encoding, the QKD-like attack is always the winning one.
\newline
The QKR channel capacity of 4-state encoding is always below 6-state.
8-state has higher capacity than 6-state
at $\qb\in[0, 0.1061]$, after which they are the same and equal to the QKD capacity.
  \item {\bf Min-entropy}.
For 4-state and 6-state, the winning attacks are as for the Shannon entropy case.
For 8-state, however, the winning attack is an ancilla attack on the key.
If capacity is computed using min-entropy loss as the measure of Eve's knowledge,
then the QKR capacity of 8-state is higher than 6-state on the range
$\qb\in[0,0.0612]$.
There is a tiny interval $\qb\in(0.0612, 0.0638)$ where 6-state outperforms 8-state;
at $\qb>0.0638$ all capacities are zero.
4-state is always worse than 6-state.
\end{itemize}
Overall, 8-state encoding requires the least privacy amplification.
\item
We notice a duality relation in the optimal POVMs for the known-plaintext attack on the key.
It turns out that the POVMs which minimise Eve's Shannon entropy are in a sense `dual' to 
the POVMs associated with the min-entropy: 
The min-entropy-POVM for plaintext $x$ is the Shannon-entropy-POVM for plaintext $1-x$.
It would be very useful if 
such dualities hold more generally. 
While there exists a simple test \cite{Holevo1973} to
check if a POVM is optimal for min-entropy,
there is no such test for Shannon entropy.
\item
As a byproduct of our analysis we find a particularly easy and insightful way
to derive the QKD capacity in a scenario where Alice adds artificial preprocessing noise.
By identifying conditional channels in Eve's mixed state we are able to 
simplify the results of \cite{SKMB2008}.
The noise-adding trick can be applied in QKR in exactly the same way as in QKD.
\end{itemize}

In Section~\ref{sec:prelim} we introduce notation, 
and briefly recap 8-state QKR.
In Section~\ref{sec:EPRandnoise} we go to the EPR version of the protocol,
apply noise symmetrisation and obtain Eve's state by purification.
Attacks on the plaintext are described in Section~\ref{sec:message},
and known-plaintext attacks on the key in Section~\ref{sec:key}.
We aggregate all the results in Section~\ref{sec:together} and we determine the QKR capacities.
Insertion of artificial noise is discussed in Section~\ref{sec:QKDwithnoise}.

\section{Preliminaries}
\label{sec:prelim}

\subsection{Notation and terminology}
\label{sec:notation}

Classical Random Variables (RVs) are denoted with capital letters, and their realisations
with lowercase letters. The probability that a RV $X$ takes value
$x$ is written as $\pr[X=x]$.
The expectation with respect to RV $X$ is denoted as 
$\EE_x f(x)=\sum_{x\in\cX}\pr[X=x]f(x)$.
The Shannon entropy of an RV $X$ is written as $\sH(X)$.
Sets are denoted in calligraphic font. 
The notation `$\log$' stands for the logarithm with base~2.
The min-entropy of $X\in\cX$ is $\Hmin(X)=-\log \max_{x\in\cX}\pr[X=x]$,
and the conditional min-entropy is
$\Hmin(X|Y)=-\log\EE_y \max_{x\in\cX}\pr[X=x|Y=y]$.
The notation $h$ stands for the binary entropy function $h(p)=p\log\fr1p+(1-p)\log\fr1{1-p}$.
Sometimes we will write $h(\{p_1,\ldots,p_n\})$ meaning $\sum_i p_i\log\fr1{p_i}$.
Bitwise XOR of binary strings is written as `$\oplus$'.
The inverse of a bit $b\in\bits$ is written as $\bar b=1-b$.

For quantum states we use Dirac notation, with the standard qubit basis
states $\ket 0$ and $\ket 1$ represented as ${1\choose 0}$ and ${0 \choose 1}$ respectively.
The Pauli matrices are denoted as $\qs_x,\qs_y,\qs_z$, and we write
$\vecsig=(\qs_x,\qs_y,\qs_z)$.
The standard basis is the eigenbasis of $\qs_z$, with $\ket0$ in the positive $z$-direction.
We write $\one$ for the identity matrix.
The notation `tr' stands for trace.
The Hermitian conjugate of an operator $A$ is written as~$A\dagg$.
When $A$ is a complicated expression, we sometimes write $(A+{\rm h.c.})$
instead of $A+A\dagg$.
The complex conjugate of $z$ is denoted as $z^*$.

We use the Positive Operator Valued Measure (POVM) formalism.
A POVM $\cM$ consists of positive semidefinite operators,
$\cM=(M_x)_{x\in\cX}$, $M_x\geq 0$, and satisfies the condition $\sum_x M_x=\one$. 
The notation $\cM(\qr)$ stands for the classical RV
resulting when $\cM$ is applied to mixed state~$\qr$.
Consider a bipartite system `AB' where the `A' part is classical, i.e.\,the state is of the form
$\qr^{\rm AB}=\EE_{x\in\cX}\ket x\bra x\otimes \qr_x$
with the $\ket x$ forming an orthonormal basis.
The min-entropy of the classical RV $X$ given part `B' of the system is \cite{KRS2009}
\be
	\Hmin(X|\qr_X)=-\log\max_\cM \EE_{x\in\cX}\tr [M_x \qr_x].
\label{defHmin}
\ee
Here $\cM$ denotes a POVM.
Let $\qL\isdef\sum_x \qr_x M_x$.
If a POVM can be found that satisfies the condition\footnote{
Ref.\,\cite{Holevo1973} specifies a second condition, namely $\qL\dagg=\qL$.
However, the hermiticity of $\qL$ already follows from the condition (\ref{Holevotest}).} 
\cite{Holevo1973}
\be
	\forall_{x\in\cX}:\; \qL-\qr_x\geq 0,
\label{Holevotest}
\ee
then there can be no better POVM (but equally good ones may exist).

For states that also depend on a classical RV $Y\in\cY$, the min-entropy of $X$ 
given the quantum state and $Y$ is
\be
	\Hmin(X|Y,\qr_X(Y))=-\log\EE_{y\in\cY}\max_\cM \EE_{x\in\cX}\tr [M_x \qr_x(y)].
\label{defHmincondY}
\ee
A simple expression can be obtained when $X$ is a binary variable.
Let $X\in\bits$.
Then
\be
	X\sim(p_0,p_1): \quad\quad
	\Hmin(X|Y,\qr_X(Y)) = 
	1-\log\left(1+\EE_y\tr\left|p_0\qr_0(y)-p_1\qr_1(y)
	\vphantom{\int^1}\right|\right).
\label{binaryHmin}
\ee
For the Shannon entropy of a classical RV given a quantum system we have
\be
	\sH(X|\qr_X)\isdef\min_{\cM}\sH(X|\cM(\qr_X)).
\ee
If the ensemble $(\qr_x)_{x\in\cX}$ has a symmetry, i.e. 
$\forall_{x\in\cX,g\in G}:\; U_g \qr_x U_g\dagg=\qr_{g(x)}$ 
for some group $G$ acting on $\cX$, and unitary representation $U$ of $G$, 
then it suffices \cite{Holevo1973} to 
consider only POVMs that obey the same symmetry,
$U_g M_x U_g\dagg=M_{g(x)}$.

\subsection{Eight-state Quantum Key Recycling}
\label{sec:QKR}

We briefly review the main properties of the 
8-state QKR scheme (``scheme \#2'' in \cite{SdV2016}).
A classical bit $g\in\bits$ is encoded into a qubit state using one of four possible bases.
The basis is labeled $b\in\{0,1,2,3\}$, and for convenience the notation
$b=2u+w$ is introduced, with $u,w\in\bits$. The labels $b$ and $(u,w)$ are used interchangeably. 
The encoding of $g$ in basis $(u,w)$ is expressed on the Bloch sphere as a unit vector
\be
	\vecn_{uwg}=\frac{(-1)^g}{\sqrt3}\left(\begin{matrix}
	(-1)^{u\phantom{+w}} \cr (-1)^{u+w} \cr (-1)^{w\phantom{+u}}
	\end{matrix}\right),
\ee
i.e. the eight corner points of a cube. The corresponding states in Hilbert space are
\be
	\ket{\qj_{uwg}}= (-1)^{gu}\left[ (-\sqrt i)^g\cos\fr\qa2\ket{g\oplus w}
	+(-1)^u(\sqrt i)^{1-g}\sin\fr\qa2\ket{\overline{g\oplus w}}\right]
\label{psiuwg}
\ee
in the $z$-basis. The angle $\qa$ is defined as $\cos\qa=1/\sqrt3$.
The four states $\ket{\qj_{uwg}}$, for fixed $g$, are the Quantum One-Time Pad (QOTP) encryptions
of $\ket{\qj_{00g}}$.

The bit error rate (BER) on the quantum channel is denoted as $\qb\in[0,\fr12]$.
The key recycling scheme makes use of a Secure Sketch $S:\bits^n\to\bits^a$,
with $a>nh(\qb)$. (Asymptotically $a$ approaches $nh(\qb)$).
Furthermore the scheme uses an extractor ${\tt Ext}:\bits^n\to\bits^\ell$
and a message-independent, key-private \cite{FehrSalvail2017}
MAC function that produces a tag of length~$\ql$.
The message is $\mu\in\bits^\ell$.
The key material shared between Alice and Bob consists of three parts:
a basis sequence $b\in\{0,1,2,3\}^n$, a MAC key $K_{\rm M}$ and a classical OTP 
$K_{\rm SS}\in\bits^a$ for protecting the secure sketch.

\noindent
\underline{Encryption}\newline
Alice performs the following steps.
Generate random $g\in\bits^n$. Compute $s=K_{\rm SS}\oplus S(g)$ 
and $z={\tt Ext}\,g$.
Compute the ciphertext $c=\mu\oplus z$ and authentication tag $T=M(K_{\rm M},g||c||s)$.
Prepare the quantum state $\ket\qJ=\bigotimes_{i=1}^n \ket{\qj_{b_i g_i}}$.
Send $\ket\qJ$, $s$, $c$, $T$.

\noindent
\underline{Decryption}\newline
(Bob gets $\ket{\qJ'}$, $s'$, $c'$, $T'$). Bob performs the following steps. 
Measure $\ket{\qJ'}$ in the b-basis. This yields $g'\in\bits^n$.
Recover $\hat g$ from $g'$ and $K_{\rm SS}\oplus s'$ (by the syndrome decoding procedure of the Secure Sketch primitive).
Compute $\hat z={\tt Ext}\,\hat g$ and 
$\hat\mu=c'\oplus\hat z$.
Accept the message $\hat\mu$
if the syndrome decoding succeeded and
$T'=M(K_{\rm M},\hat g||c'||s')$.
Communicate Accept/Reject to Alice.

\noindent
\underline{Key update}\newline
Alice and Bob perform the following actions.
If Bob Accepts, replace $K_{\rm SS}$.
If Bob Rejects, replace $K_{\rm SS}$ and compute the updated key $b'$ as a function of $b$ and $n$ fresh secret bits.

\vskip2mm

In case of Bob accepting the transmission, an $\ell$-bit message has been communicated
while only $a\approx nh(\qb)$ bits of key material have been spent.\footnote{
``Scheme \#3'' in \cite{SdV2016} greatly reduces the key material expenditure.
}
The aim of the current paper is to find out how large $\ell$ is allowed to be as a function of the noise
parameter~$\qb$.

\section{EPR formulation, noise symmetrisation, and purification}
\label{sec:EPRandnoise}

Apart from QKR employing the 8-state (QOTP) encoding
as described above,
we also investigate 4-state (BB84) and 6-state conjugate coding.
For the security analysis of qubit-based QKR we piggyback on 
(i)
proof techniques \cite{Renner2007} that use e.g.\;quantum de Finetti \cite{CKMR2007}
to reduce the analysis to individual-qubit attacks;
(ii)
the proof technique for qubit-based QKR introduced in \cite{FehrSalvail2017},
which can directly be applied to the scheme of \cite{SdV2016} provided that
correct values are known for the required amount of privacy amplification
as a function of the noise parameter~$\qb$.

We study optimal attacks against individual qubits, making use of 
the standard Shor-Preskill technique \cite{SP2000} and
the noise symmetrisation technique introduced by \cite{RGK2005}.

\subsection{EPR version of the QKR protocol}
\label{sec:EPR}

We follow the standard Shor-Preskill technique \cite{SP2000} and
re-formulate the QKR protocol (Section~\ref{sec:QKR}) using EPR pairs.
The step where Alice prepares the state $\ket\qJ$ and sends it to Bob
is replaced by the following procedure.

Alice prepares a two-qubit singlet state. She keeps one qubit (`A') and sends the other qubit (`B') to Bob.
Eve is allowed to manipulate the whole `AB' system\footnote{
Note that this attacker model gives Eve more power than she can actually have in real life.
Realistically, she would be able to manipulate only the `B' subsystem.
} in any way, including coupling to ancillas.
Then Alice and Bob perform their projective measurements in the correct basis
(basis $b_i$ for the $i$'th bit). Let the outcome of Alice's measurement be $x\in\bits$,
and Bob's outcome $y\in\bits$.
Alice sends $e=x\oplus g$ to Bob. Bob computes $\hat g=\bar y\oplus e$, which 
is guaranteed to equal~$g$ if Eve has done nothing ($\qb=0$).\footnote{
In the singlet state the $x$ and $y$ are anti-correlated, i.e. $y=\bar x$.
}
Security of this EPR-version of the protocol implies security of the original protocol.

Note that the above description is agnostic of the number of bases used in the encoding.
We will use the notation $\cB$ to denote the set of bases in an encoding scheme.
For 4-state encoding we write $\cB=\{0,1\}$,
and the states are the spin states $\ket{\pm z}$ (at $b=0$) and $\ket{\pm x}$ (at $b=1$).
For 6-state we write $\cB=\{1,2,3\}$, with spin states $\ket{\pm x}$ (at $b=1$),
$\ket{\pm y}$ (at $b=2$) and $\ket{\pm z}$ (at $b=3$).
For 8-state we have $\cB=\{00,01,10,11\}$, and the states are defined in~(\ref{psiuwg}).
The number of bases is $|\cB|$.

\subsection{Noise symmetrisation}

After Eve's interference,
the bipartite system held by Alice and Bob is no longer a pure singlet state
but a general mixed state $\qr^{\rm AB}$.
As the singlet state is invariant under unitary transformations of the form
$\qr^{\rm AB}\mapsto U\otimes U\qr^{\rm AB}U\dagg\otimes U\dagg$
(where $U$ acts on a single qubit), 
Alice and Bob are `allowed' to perform the following sequence of actions.

\vskip1mm
\underline{Preparation phase, before the protocol}\\
Alice and Bob agree on a single basis $b^*\in\cB$. 

\vskip1mm
\underline{During the protocol}\\
For each bit, just before they execute their measurement
\begin{itemize}
\item
Alice and Bob publicly draw a random number $\qg\in\{0,1,2,3\}$.
\item
They both apply to their own qubit the Pauli operator $\qs_\qg$, defined with respect to the
$b^*$ basis. Here $\qs_0$ is the identity matrix.
\item
They forget $\qg$.
\end{itemize}
These actions have no effect on the original state (the desired singlet) but they dramatically simplify
the noise in $\qr^{\rm AB}$.

\begin{lemma}
\label{lemma:noisesymm}
Consider 6-state or 8-state encoding.
Let $\ket{\qJ^\pm}=\frac{\ket{01}_*\pm\ket{10}_*}{\sqrt2}$
and $\ket{\qF^\pm}=\frac{\ket{00}_*\pm\ket{11}_*}{\sqrt2}$ denote the Bell basis states
with respect to the $b^*$ basis.
Let Eve introduce a bit error rate of {\em exactly} $\qb$ between Alice and Bob's measurement results.
Then the mixed state of the `AB' system after the above described symmetrisation procedure is given by
\be
	\tilde\qr^{\rm AB}=(1-\frac32\qb)\ket{\qJ^-}\bra{\qJ^-}+\frac\qb2\left(
	\ket{\qF^-}\bra{\qF^-} + \ket{\qJ^+}\bra{\qJ^+} +\ket{\qF^+}\bra{\qF^+}
	\vphantom{\int}\right).
\label{noiseform}
\ee
\end{lemma}
\underline{Proof}:
In \cite{RennerThesis} it was shown that the AB state reduces to the form
$\tilde\qr=\ql_0\ket{\qJ^-}\bra{\qJ^-}+\ql_1\ket{\qF^-}\bra{\qF^-} + \ql_2\ket{\qJ^+}\bra{\qJ^+} +\ql_3\ket{\qF^+}\bra{\qF^+}$,
with $\ql_0+\ql_1+\ql_2+\ql_3=1$.
We impose the constraint 
$(\ket{\qj_{bg}}\otimes \ket{\qj_{bg}})\dagg\tilde\qr\ket{\qj_{bg}}\otimes \ket{\qj_{bg}}=\qb/2$ 
for all $b\in\cB$, $g\in\bits$.\footnote{
From the above constraints and $\tr\tilde\qr=1$ it follows that
$(\ket{\qj_{bg}}\otimes \ket{\qj_{b\bar g}})\dagg\tilde\qr\ket{\qj_{bg}}\otimes \ket{\qj_{b\bar g}}=\frac{1-\qb}2$.  
}
For the 6-state case it was shown in \cite{RGK2005} that these constraints yield (\ref{noiseform}).
We next study the 8-state case. Taking $b=b^*$, the above constraints yield
$\fr12\ql_2+\fr12\ql_3=\fr\qb2$.
The case $b\neq b^*$ is more complicated.
Without loss of generality we take $b^*=00$. 
Then the $b=01$ and $b=11$ constraints each give, after some algebra,
$\fr1{18}(7\ql_1+8\ql_2+3\ql_3)=\fr\qb2$.
The $b=10$ constraint gives $\fr1{18}(\ql_1+8\ql_2+9\ql_3)=\fr\qb2$.
Solving for the $\ql$-parameters finally yields $\ql_1=\ql_2=\ql_3=\fr\qb2$.
\hfill$\square$

Note that setting $b^*\in\cB$ is important:
if the Pauli operators $\qs_\qg\otimes \qs_\qg$ are chosen with respect to a different basis,
then Lemma~\ref{lemma:noisesymm} does not necessarily hold.

Also note that Lemma~\ref{lemma:noisesymm} usually does not hold for 4-state (BB84) conjugate coding.
4-state encoding has fewer noise-related constraints, and hence Eve has more freedom.
However, one can imagine a protocol variant where Alice and Bob spend some extra key material\footnote{
This key has to be refreshed every time, otherwise Eve may find out which positions are test positions.
}
in order to agree on qubit positions which they sacrifice for noise testing purposes.
With Lemma~\ref{lemma:noisesymm} holding for 4-state too,
we can now treat all three encoding methods on an equal footing.
We will see in Section~\ref{sec:together} that even with this advantage given to Alice and Bob for 4-state,
the 4-state encoding still performs worst.

\subsection{Purification}
\label{sec:purification}

The $\tilde\qr^{\rm AB}$ can be purified as follows, under the worst-case assumption that
all noise is caused by Eve. 
Denoting Eve's four-dimensional subsystem as `E', with orthonormal basis $\ket{m_i}$, we can write
\be
	\ket{\qJ^{\rm ABE}}=\sqrt{1-\fr32\qb}\ket{\qJ^-}\otimes\ket{m_0}
	+\sqrt{\fr\qb2}\left(
	-\ket{\qF^- }\otimes\ket{m_1} + i\ket{\qJ^+ }\otimes\ket{m_2} + \ket{\qF^+ }\otimes\ket{m_3}
	\vphantom{\int}\right).
\label{purifiedABE}
\ee
Alice and Bob know in which basis to measure. 
They both do a projective measurement on their own subsystem.
They measure the spin component in the direction
$\vecv=(v_x,v_y,v_z)=(\sin\qy\cos\qf,\sin\qy\sin\qf,\cos\qy)$.
The eigenstates of this measurement are
$\ket{\vecv}=e^{-i\qf/2}\cos\fr\qy2\ket0+e^{i\qf/2}\sin\fr\qy2\ket1$ 
(with eigenvalue `0')
and
$\ket{\overline \vecv}=-e^{-i\qf/2}\sin\fr\qy2\ket0+e^{i\qf/2}\cos\fr\qy2\ket1$
(with eigenvalue `1').

We rewrite the state (\ref{purifiedABE}) using $\ket \vecv, \ket{\overline \vecv}$
as the basis of the A and B subsystem,
\bea
	\ket{\qJ^{\rm ABE}}&=&
	\sqrt{\fr{1-\qb}2}\ket{\vecv\overline \vecv}\otimes\ket{E^\vecv_{01}}
	-\sqrt{\fr{1-\qb}2}\ket{\overline \vecv\vecv}\otimes\ket{E^\vecv_{10}}
	+\sqrt{\fr\qb2}\ket{\vecv\vecv}\otimes\ket{E^\vecv_{00}}
	-\sqrt{\fr\qb2}\ket{\overline \vecv\,\overline \vecv}\otimes\ket{E^\vecv_{11}}
	\nn\\
	\ket{E^\vecv_{01}} &=& \frac1{\sqrt{1-\qb}}\left[
	\sqrt{1-\fr32\qb}\ket{m_0}+\sqrt{\fr\qb2}\left(v_x\ket{m_1}+v_y\ket{m_2}+v_z\ket{m_3}
	\right)\right]
	\nn\\
	\ket{E^\vecv_{10}} &=& \frac{1}{\sqrt{1-\qb}}\left[
	\sqrt{1-\fr32\qb}\ket{m_0}-\sqrt{\fr\qb2}\left(v_x\ket{m_1}+v_y\ket{m_2}+v_z\ket{m_3}
	\right)\right]
	\nn\\
	\ket{E^\vecv_{00}} &=& \frac1{\sqrt{2(1-v_z^2)}}\left[(-v_x v_z-iv_y)\ket{m_1}
	+(-v_y v_z+iv_x)\ket{m_2}+(1-v_z^2)\ket{m_3}\right]
	\nn\\
	\ket{E^\vecv_{11}} &=&
	\frac{1}{\sqrt{2(1-v_z^2)}}\left[(-v_x v_z+iv_y)\ket{m_1}+(-v_y v_z-iv_x)\ket{m_2}+(1-v_z^2)\ket{m_3}\right].
\label{vvbasis}
\eea
A number of things are worth noting about this representation of the purification.
\begin{itemize}
\item
With probability $1-\qb$, Alice and Bob's measurement outcomes are opposite.
With probability $\qb$ they are equal.
\item
$\ket{E^\vecv_{10}}=\ket{E^{-\vecv}_{01}}$ and
$\ket{E^\vecv_{11}}=\ket{E^{-\vecv}_{00}}$.
Furthermore $\inprod{E^\vecv_{00}}{E^\vecv_{11}}=0$, and
$\ket{E^\vecv_{00}}$, $\ket{E^\vecv_{11}}$ span a subspace orthogonal to
$\ket{E^\vecv_{01}}$, $\ket{E^\vecv_{10}}$. 
Furthermore, 
$\inprod{E^\vecv_{01}}{E^\vecv_{10}}=\frac{1-2\qb}{1-\qb}$.
This structure makes it particularly easy to analyse QKD. See Section~\ref{sec:M2Shannon}.
\item
$|\frac{-v_x v_z-iv_y}{\sqrt{1-v_z^2}}|^2=1-v_x^2$
and
$|\frac{-v_y v_z+iv_x}{\sqrt{1-v_z^2}}|^2=1-v_y^2$.
\end{itemize}

In the analysis of QKD schemes, it suffices to express (\ref{vvbasis}) only for a single choice of~$\vecv$,
because the basis is eventually revealed to Eve.
In QKR the basis is not revealed. 
In our treatment of known plaintext attacks (Section~\ref{sec:key})
we will need to evaluate (\ref{vvbasis}) for different bases.

\subsection{Eve's mixed state} 
\label{sec:Evestate}

After Alice and Bob have performed their measurement,
Eve possesses one of the $4|\cB|$ pure states $\qr^{\vecv(b)}_{xy}$, with $x,y\in\bits$, $b\in\cB$
\be
	\qr^\vecv_{xy}\isdef \ket{E^\vecv_{xy}}\bra{E^\vecv_{xy}},
\ee
coupled to the unknown (to her) classical random variables $B,X,Y$.
The whole system of $B,X,Y$ and E can be represented as a four-part
system in the following mixed state,
\be
	\qO^{BXYE}=\frac1{|\cB|}\sum_{b\in\cB} \EE_{x\in\bits}\EE_{y|x} \ket b\bra b\otimes
	\ket x\bra x\otimes \ket y\bra y \otimes \qr^{\vecv(b)}_{xy}.
\ee
At given $x$, the probability of $y\neq x$ is $1-\qb$.
(Before the introduction of noise, the $x$ and $y$ were perfectly anti-correlated.)

In Section~\ref{sec:key} we will study known plaintext attacks, i.e. Eve knows $g$ and wants to learn the basis~$b$.
If Eve knows that $x=0$, then she has to distinguish between the following $|\cB|$ states, 
\be
	\qz_b\isdef(1-\qb)\qr^{\vecv(b)}_{01}+\qb \qr^{\vecv(b)}_{00},
	\quad\quad b\in\cB.
\label{defzeta}
\ee
The case $x=1$ will not be treated separately as it is analogous to $x=0$.

\section{Security of the message}
\label{sec:message}

\subsection{Attacks targeting the message}

We consider attacks by which Eve tries to 
gain information about Alice's plaintext $x$.
\begin{itemize}
\item[{\bf M1}] 
Eve steals one whole transmission $\ket\qJ$ and performs a measurement. 
(No matter what Eve sends to Bob, Bob rejects with overwhelming probability.)
\item[{\bf M2}]
Eve couples each qubit individually to an ancilla, and transfers information into the ancilla
in such a way that the bit error rate is exactly~$\qb$. 
She does this for $N$ transmissions ($N\gg 1$) before finally performing a measurement on her ancillas.
\end{itemize}

Attack M1 is the worst case scenario given that Bob does not accept.
M2 is the worst case given that Bob accepts $N$ times in a row.

Attack M1 has no effect against 8-state encoding (since it is a QOTP), but is important in the case
of 4-state and 6-state encoding. 
Below we briefly recap the results of \cite{SdV2016}. 
In Section~\ref{sec:M2} we will see that the analysis of M2
reduces to the analysis of QKD.

\subsection{Attack M1 on 4-state encoding}
\label{sec:M14}

Eve intercepts the whole $n$-qubit state $\ket\qJ$
and immediately does a measurement.
She subjects each qubit $i$ individually to the spin measurement
$(\qs_x+\qs_z)/\sqrt2$. 
The probability distribution
of $X_i$ given the outcome always consists of the numbers
$(\cos\fr\pi8)^2$ and $(\sin\fr\pi8)^2$.
In terms of Shannon entropy this corresponds to the following mutual information per qubit,
\be
	I^{\rm M1,4state}_{\rm AE}=1-h([\sin\fr\pi8]^2)\approx 0.399.
\ee
The min-entropy loss per qubit is
\be
	\tri\Hmin^{\rm M1,4state}=1-\log\frac1{(\cos\fr\pi8)^2}\approx 0.772.
\ee

\subsection{Attack M1 on 6-state encoding}
\label{sec:M16}

Eve's spin measurement is $(\qs_x+\qs_y+\qs_z)/\sqrt3$.
The probability distribution for $X_i$ given the outcome always consists of the numbers
$(\cos\fr\qa2)^2$ and $(\sin\fr\qa2)^2$. This yields
\bea
	I^{\rm M1,6state}_{\rm AE} &=& 1-h([\sin\fr\qa2]^2)\approx 0.256
	\\
	\tri\Hmin^{\rm M1,6state}&=&1-\log\frac1{(\cos\fr\qa2)^2}\approx 0.658.
\eea

\subsection{Attack M2: All Your Basis Are Belong To Us.}
\label{sec:M2}

Attack M2 is effective because Eve is attacking $N$ qubits that are encrypted {\em with
the same key~$b$}.
Eve collects $N$ ancillas containing partial information about the message bits;
these message bits are protected by a total of $\log|\cB|$ key bits.
Hence, for large $N$ the key $b$ offers essentially no protection of the information drawn into the ancillas. 
(On the other hand, the key prevents Eve
from absorbing full information into her ancillas. And the key itself does not become known to Eve.)

\begin{lemma}
\label{lemma:equivQKD}
Let Alice and Bob take fresh keys and then run the EPR version of the
QKR protocol $N$ times, with Bob Accepting each time.
Let $X_i^{(j)}$, with $j\in\{1,\ldots,N\}$, be
Alice's measurement result in qubit position $i\in\{1,\ldots,n\}$
in the $j$'th run of the protocol
and $B_i$ the basis key used to encode all the $X_i^{(j)}$.
Let $E_i^{(j)}$ denote Eve's corresponding ancilla system, 
created without knowledge of $B_i$. 
Then
\be
	\frac1N \sH(X_i^{(1)},\ldots,X_i^{(N)}|\; E_i^{(1)},\ldots,E_i^{(N)}) \geq 
	\sH(X_i^{(j)}| B_i E_i^{(j)})\quad\quad j\;{\rm arbitrary}.
\label{AYBABTU}
\ee
\end{lemma}
\underline{Proof}:
Let $\cM$ denote a POVM.
We have $\sH(\vecX_i|\vecE_i)=$ $\min_\cM\sH(\vecX_i|\cM(\vecE_i))$ $\geq \min_\cM\sH(\vecX_i|B_i \cM(\vecE_i))$
$=N\min_\cM\sH(X_i^{(j)}|B_i \cM(E_i^{(j)}))$ 
$=N\sH(X_i^{(j)}| B_i E_i^{(j)})$ for arbitrary~$j$.
\hfill$\square$

For $N\gg 1$ the bound is tight.
The left hand side of (\ref{AYBABTU}) is the leakage per qubit.
The right hand side
is precisely the quantity that determines the security of QKD:
the uncertainty about $X$ given a noise-constrained ancilla
and the basis $B$ revealed to Eve {\em after she has created the ancilla states}. 

Lemma~\ref{lemma:equivQKD} allows us to obtain a tight lower bound on the QKR capacity,
namely the QKD capacity, whenever M2 is the dominant attack.

\subsubsection{QKD, Shannon entropy}
\label{sec:M2Shannon}

The computation of $\sH(X|BE)$ for BB84 and 6-state (or more) QKD is well known.
Here we combine the two standard approaches:
(i) the simplest possible description of the noise, i.e. noise symmetrisation,
(ii) specifying optimal measurements instead of bounds based on von Neumann entropy.
The results are of course not new, but we present the matter in a particularly clean way
which helps when protocol embellishments are considered (e.g. addition of artificial noise,
see Section~\ref{sec:QKDwithnoise}).

\underline{Informal treatment}\\
Eve knows $\vecv$.
Eve does a projective measurement $\ket{E^\vecv_{00}}\bra{E^\vecv_{00}}+\ket{E^\vecv_{11}}\bra{E^\vecv_{11}}$.
This measurement does not destroy any information.
With probability $\qb$ the outcome is `1'; next Eve can perfectly distinguish between the orthogonal 
states $\ket{E^\vecv_{00}}$, $\ket{E^\vecv_{11}}$ and hence learns $X$ with 100\% accuracy.
With probability $1-\qb$ the outcome is `0'; now
Eve has to handle the trickier task of distinguishing between the non-orthogonal $\ket{E^\vecv_{01}}$
and $\ket{E^\vecv_{10}}$, which have inner product
$c\isdef\inprod{E^\vecv_{01}}{E^\vecv_{10}}=\frac{1-2\qb}{1-\qb}$.
This is done optimally using a projective measurement in the following orthonormal basis,
\bea
	\ket{\mu_{01}}&=& \qg_+\ket{E^\vecv_{01}}+\qg_-\ket{E^\vecv_{10}}
	\nn\\
	\ket{\mu_{10}}&=& \qg_+\ket{E^\vecv_{10}}+\qg_-\ket{E^\vecv_{01}}
	\nn\\
	\qg_\pm &=& \frac1{2\sqrt{1+c}}\pm\frac1{2\sqrt{1-c}}
\eea  
and has error probability 
\be
	p_\qb=|\inprod{E^\vecv_{01}}{\mu_{10}}|^2=|\inprod{E^\vecv_{10}}{\mu_{01}}|^2
	=\fr12-\fr12\sqrt{1-c^2}
	=\fr12-(1-\qb)^{-1}\sqrt{\fr\qb2(1-\fr32\qb)}.
\label{defpbeta}
\ee
The channel capacity from Alice to Eve is 
\be
	I_{\rm AE}(\qb)=\qb\cdot[1-h(0)]+(1-\qb)[1-h(p_\qb)].
\label{IAEQKD}
\ee
The secrecy capacity is 
\be
	C(\qb)=I_{\rm AB}(\qb)-I_{\rm AE}(\qb)=1-h(\qb)-I_{\rm AE}(\qb).
\label{capQKD}
\ee

\underline{Formal treatment}\\
Eve has to guess $X$ from a state $\qr^\vecv_{XY}=\ket{E^\vecv_{XY}}\bra{E^\vecv_{XY}}$.
We write $Y=\bar X\oplus R$, with $R\in\bits$ the noise. Eve does not know~$R$.
Let $\cQ=(Q_{x})_{x\in\bits}$ be a POVM applied by Eve, and let $\cQ(\qr^\vecv_{XY})\in\bits$ be the outcome of the measurement.
The main quantity to compute is 
\bea
	\sH(X|\qr^\vecv_{X,\bar X\oplus R}) &=& \min_\cQ \sH(X|\cQ(\qr^\vecv_{X,\bar X\oplus R}))
	=\min_\cQ \EE_r\sH(X|\cQ(\qr^\vecv_{X,\bar X\oplus r}))
	\nn\\ &=&
	\min_\cQ \left[(1-\qb)\sH(X|\cQ(\qr^\vecv_{X\overline X}))+\qb\sH(X|\cQ(\qr^\vecv_{XX}))
	\vphantom{\int^1}\right].
\eea
The optimal POVM is given by 
$Q_0=\ket{E^\vecv_{00}}\bra{E^\vecv_{00}}+\ket{\mu_{01}}\bra{\mu_{01}}$,
$Q_1=\ket{E^\vecv_{11}}\bra{E^\vecv_{11}}+\ket{\mu_{10}}\bra{\mu_{10}}$.
This is equivalent to the two-step procedure detailed in the informal treatment above, and yields
\be
	\sH(X|\qr^\vecv_{XY}) = (1-\qb)h(p_\qb) + \qb\cdot 0.
\ee
Eve's knowledge about $X$ is $I_{\rm AE}=\sH(X)-\sH(X|\qr^\vecv_{XY})$, which precisely equals (\ref{IAEQKD}).

\subsubsection{QKD, min-entropy}
\label{sec:M2Hmin}

Expressed as min-entropy loss, Eve's knowledge is
$\Hmin(X)-\Hmin(X|\qr^\vecv_{X,\bar X\oplus R})$
for known $\vecv$ and unknown noise $R\in\bits$.
We have
\bea
	\Hmin(X|\qr^\vecv_{X,\overline X\oplus R}) &=&
	-\log p_{\rm guess}(X|\cQ(\EE_r\qr^\vecv_{X,\overline X\oplus r}))
	\nn\\ &=&
	-\log\EE_r p_{\rm guess}(X|\cQ(\qr^\vecv_{X,\overline X\oplus r}))
	\nn\\ &=&
	-\log\left[ \qb p_{\rm guess}(X|\cQ(\qr^\vecv_{XX}))
	+ (1-\qb)p_{\rm guess}(X|\cQ(\qr^\vecv_{X\overline X})) 
	\right]
	\nn\\ &=&
	-\log\left[ \qb\cdot 1+(1-\qb)(1-p_\qb) \right]
	\nn\\ &=&
	\Hmin(X)-\log[1+\sqrt2\sqrt{\qb(1-\fr32\qb)}+\qb].
\eea

\section{Security of the key}
\label{sec:key}

\subsection{Known plaintext attacks on the key}
\label{sec:kayattacks}

We have to take into account the possibility that
Eve knows the plaintext~$\mu$.
Then $\qJ$ may give Eve information on the (basis) key~$b$.
We focus on attacks that lead Bob to Accept. 
(A Reject causes Alice and Bob to refresh their keys.)
We look at the two types of attack available to Eve,
\begin{itemize}
\item[{\bf K1}]
Eve intercepts a fraction $3\qb$ of the qubits, does a measurement on them, and
sends the resulting states on to Bob.
\item[{\bf K2}]
Eve lets every qubit individually interact with an ancilla.
She forwards the qubits to Bob.
\end{itemize}

In attack K1 Eve receives a state
\be
	\qo_{Bx}=\ket{\qj_{Bx}}\bra{\qj_{Bx}}
\ee
for known $x$ and unknown~$B$.
For attack K2 Eve's view is the mixed state $\qz_B$
as defined in (\ref{defzeta}), for unknown $B$.

\begin{lemma}
\label{lemma:KShannon}
The Shannon entropy of $B$ given $\qz_B$ can be written as
\be
	\sH(B|\qz_B)=\log |\cB| -\max_\cM\left[
	h(\{\tr M_m \frac{\sum_b \qz_b}{|\cB|} \}_{m\in\cB})
	-\frac1{|\cB|}\sum_{b\in\cB} h(\{\tr M_m \qz_b\}_{m\in\cB})
	\right]
\label{ShannonK2general}
\ee
where $\max_\cM$ is maximisation over POVMs $(M_m)_{m\in\cB}$.
If we impose the symmetry relations
$\forall_{b\in\cB}:\; \tr M_b\qz_b=\pok$
and $\forall_{m,b\in\cB, m\neq b}:\; \tr M_m\qz_b=\frac{1-\pok}{|\cB|-1}$ then 
the expression for the entropy reduces to
\be
	\sH(B|\qz_B)=\min_{{\rm symmetric}\,\cM}\left[h(\pok)+(1-\pok)\log(|\cB|-1)
	\vphantom{M^M}\right].
\label{K2hshort}
\ee
\end{lemma}
\underline{Proof}:
Let $\cM(\qz_B)$ be the classical random variable describing the outcome of the POVM measurement
$\cM$ on state $\qz_B$.
We have $\sH(B|\qz_B)=\min_\cM \sH(B|\cM(\qz_B))$, with
$\sH(B|\cM(\qz_B))=\sum_m \pr[\cM(\qz_B)=m]\sH(B|\cM(\qz_B)=m)$.
We write $\pr[B=b|\cM(\qz_B)=m]=\frac1{|\cB|}[\tr M_m\qz_b]/\pr[\cM(\qz_B)=m]$
and $\pr[\cM(\qz_B)=m]=\frac1{|\cB|}\sum_b\tr M_m\qz_b$.
After some manipulation (\ref{ShannonK2general}) follows.
In the first $h(\cdots)$ of (\ref{ShannonK2general}) we then write
$\fr1{|\cB|}\sum_b\tr\qz_b M_m=\fr1{|\cB|}[\pok +(|\cB|-1)\fr{1-\pok}{|\cB|-1}]=\fr1{|\cB|}$.
The $h(\fr1{|\cB|})$ cancels the $\log|\cB|$.
The second $h(\cdots)$ in (\ref{ShannonK2general}) is the same for all $b\in\cB$,
namely $h(\{\pok,\frac{1-\pok}{|\cB|-1},\ldots,\frac{1-\pok}{|\cB|-1}\})$
$=-\pok\log\pok-(|\cB|-1)\cdot\frac{1-\pok}{|\cB|-1}\log\frac{1-\pok}{|\cB|-1}$
$=h(\pok)+(1-\pok)\log(|\cB|-1)$.
\hfill$\square$

\subsection{Attack K1, 4-state}
\label{sec:K14}

Eve scrutinises $\qo_{Bx}$.
If $x=0$ then the state is either the $+x$ or $+z$ spin state.
If $x=1$ then the state is either $-x$ or $-z$.
In both cases, the optimal way to distinguish between the states
is to measure the spin $(\qs_x-\qs_z)/\sqrt2$.
Given the measurement outcome, the probabilities for the two key values
are $(\cos\fr\pi8)^2$ and $(\sin\fr\pi8)^2$. This holds for $x=0$ as well as $x=1$.
Eve's knowledge about $B$ is
\bea
	\sH(B)-\sH(B|X,\qo_{BX}) &=& 1-h([\sin\fr\pi8]^2)\approx 0.399
\label{K14Shannon}
	\\
	\Hmin(B)-\Hmin(B|X,\qo_{BX}) &=& 1-\log\frac1{(\cos\fr\pi8)^2}\approx 0.772.
\label{K14Hmin}
\eea
The effect on the whole $n$-bit string is obtained by multiplying
(\ref{K14Shannon},\ref{K14Hmin}) times $3\qb n$.

\subsection{Attack K1, 6-state}
\label{sec:K16}

Consider $x=0$. (The analysis for $x=1$ is analogous).
Eve has to distinguish between the spin states
$+x$, $+y$, $+z$ using a POVM $\cM=(M_b)_{b\in\{1,2,3\}}$.
For the min-entropy the best POVM is given by
$M_b=\fr13\one-\fr13\vecn_b\cdot\vecsig$, with
$\vecn_1=(-2,1,1)^{\rm T}/\sqrt6$,
$\vecn_2=(1,-2,1)^{\rm T}/\sqrt6$,
$\vecn_3=(1,1,-2)^{\rm T}/\sqrt6$.
It yields the following probability distribution for $B$:
$\{\fr13+\fr2{3\sqrt6}$, $\fr13-\fr1{3\sqrt6}$, $\fr13-\fr1{3\sqrt6}\}$.
\be
	\Hmin(B)-\Hmin(B|X,\qo_{BX}) = \log 3+\log(\fr13+\fr2{3\sqrt6})\approx 0.861.
\label{K16Hmin}
\ee
For the Shannon entropy the best POVM is of the same form as above but with
$\vecn_b\to -\vecn_b$. The probability distribution for $B$ is
$\{\fr13+\fr1{3\sqrt6}$, $\fr13+\fr1{3\sqrt6}$, $\fr13-\fr2{3\sqrt6}\}$. 
\be
	\sH(B)-\sH(B|X,\qo_{BX})
	=\log3-h(\{\fr13+\fr1{3\sqrt6}, \fr13+\fr1{3\sqrt6}, \fr13-\fr2{3\sqrt6}\})
	\approx 0.314.
\label{K16Shannon}
\ee
The effect on the whole $n$-bit string is obtained by multiplying
(\ref{K16Hmin},\ref{K16Shannon}) times $3\qb n$.

\subsection{Attack K1, 8-state}
\label{sec:K18}

Consider $x=0$. (The analysis for $x=1$ is analogous).
Eve has to distinguish between the four states $\ket{\qj_{b0}}$
with a POVM $\cM=(M_b)_{b\in\cB}$.
For the min-entropy the optimal POVM is $M_b=\fr12\ket{\qj_{b0}}\bra{\qj_{b0}}$,
yielding probability distribution $\{\fr12,\fr16,\fr16,\fr16\}$.
For the Shannon entropy the optimum is
$M_b=\fr12\ket{\qj_{b1}}\bra{\qj_{b1}}$,
yielding distribution $\{0,\fr13,\fr13,\fr13\}$.
\bea
	\Hmin(B)-\Hmin(B|X,\qo_{BX}) &=& 2-1 =1
\label{K18Hmin}
	\\
	\sH(B)-\sH(B|X,\qo_{BX}) &=& 2-\log 3\approx 0.415.
\label{K18Shannon}
\eea
The effect on the whole $n$-bit string is obtained by multiplying
(\ref{K18Hmin},\ref{K18Shannon}) times $3\qb n$.

\subsection{Attack K2, 4-state}
\label{sec:K24}

Eve has to distinguish between $B=0$ ($z$-basis) and $B=1$ ($x$-basis)
by inspecting her ancilla state~$\qz_B$.

\begin{theorem}
\label{th:K24}
In the case of 4-state encoding, the min-entropy of the basis $B$ given the mixed state $\qz_B$
is
\be
	\Hmin(B|\qz_B)=\Hmin(B)-\log(1+\sqrt{\qb(1-\fr32\qb)}+\frac\qb{\sqrt2}).
\label{POVM4Hmin}
\ee
The corresponding POVM $\cM=(M_b)_{b\in\bits}$ is given by
\bea
	M_0=\ket{\qg_1}\bra{\qg_1}+\ket{\qg_2}\bra{\qg_2}
	&;&
	M_1=\ket{\qg_3}\bra{\qg_3}+\ket{\qg_4}\bra{\qg_4}
	\\
	\ket{\qg_1}= \frac{\ket{m_0}}{\sqrt2}+\frac{\ket{m_3}-\ket{m_1}}2
	&;&
	\ket{\qg_3}= \frac{\ket{m_0}}{\sqrt2}-\frac{\ket{m_3}-\ket{m_1}}2
	\nn\\
	\ket{\qg_2}=\frac{\ket{m_2}}{\sqrt2}+i\frac{\ket{m_1}+\ket{m_3}}2
	&;&
	\ket{\qg_4}=\frac{\ket{m_2}}{\sqrt2}-i\frac{\ket{m_1}+\ket{m_3}}2.
\eea
\end{theorem}

\underline{Proof}:\\
\bea
	\ket{E^{(0,0,1)}_{01}} =
	\frac{\sqrt{1-\fr32\qb}\ket{m_0}+\sqrt{\fr\qb2}\ket{m_3}}
	{\sqrt{1-\qb}}
	&;&
	\ket{E^{(1,0,0)}_{01}} =
	\frac{\sqrt{1-\fr32\qb}\ket{m_0}+\sqrt{\fr\qb2}\ket{m_1}}
	{\sqrt{1-\qb}}
	\nn\\
	\ket{E^{(0,0,1)}_{00}}\propto \frac{\ket{m_1}-i\ket{m_2}}{\sqrt2}
	&;&
	\ket{E^{(1,0,0)}_{00}}=\frac{i\ket{m_2}+\ket{m_3}}{\sqrt2}
\eea
\be
	\qz_{0}-\qz_{1}=\sqrt{\qb(1-\fr32\qb)}\left[\ket{m_0}\frac{\bra{m_3}-\bra{m_1}}{\sqrt2} +{\rm h.c.}\right]
	+\frac\qb{\sqrt2}\left[
	-i\ket{m_2}\frac{\bra{m_1}+\bra{m_3}}{\sqrt2} +{\rm h.c.}
	\right].
\ee
The two expressions between square brackets act on orthogonal two-dimensional subspaces and both have the form
of a Pauli operator. It directly follows that the eigenvalues are $\pm\sqrt{\qb(1-\fr32\qb)}$
and $\pm \qb/\sqrt2$.
Finally we apply (\ref{binaryHmin}) with $p_0=p_1=\fr12$.
\hfill$\square$

\begin{theorem}
\label{th:K24Shannon}
In the case of 4-state encoding, the Shannon entropy of the basis $B$ given the mixed state $\qz_B$
is 
\be
	\sH(B|\qz_B)=h(\frac12+\frac12\sqrt{\qb(1-\fr32\qb)}+\frac{\qb}{2\sqrt2}).
\label{H4conj}
\ee
\end{theorem}
\underline{Proof}:
For binary $B$, the POVM associated with the min-entropy maximises $\tr M_0(\qz_0-\qz_1)$
(see Section~\ref{sec:notation}). If we impose the symmetry $\tr M_0\qz_1=\tr M_1\qz_0$
then this expression becomes $\tr M_0\qz_0-(1-\tr M_0\qz_0)=2\tr M_0\qz_0-1$.
(Imposing this symmetry is allowed, see Section~\ref{sec:notation}).
Hence the optimisation in the min-entropy-POVM is the same as the optimisation in 
the Shannon-POVM, and we conclude that the POVM associated with the min-entropy also
minimises the Shannon entropy.
Applying the POVM from Theorem~\ref{th:K24} to (\ref{K2hshort}) yields (\ref{H4conj}).
\hfill$\square$

\subsection{Attack K2, 6-state}
\label{sec:K26}

Eve has to distinguish between $B=1$ ($x$-basis), $B=2$ ($y$-basis),
and $B=3$ ($z$-basis).
We define the permutation matrix $S$ as
\be
	S\isdef \ket{m_0}\bra{m_0}+\ket{m_2}\bra{m_1}+\ket{m_3}\bra{m_2}+\ket{m_1}\bra{m_3}.
\ee

\begin{theorem}
\label{th:K26}
In the case of 6-state encoding, the min-entropy of the basis $B$ given the mixed state $\qz_B$
is
\be
	\Hmin(B|\qz_B)=\Hmin(B)
	-\log\left(1+\frac{2\sqrt2}{\sqrt3}\sqrt{\qb(1-\qb)}\right).
\ee
The associated POVM is
\bea
	M_3 &=&  \frac{3-4\qb}{3(1-\qb)}\ket{q}\bra{q}+\frac1{3(1-\qb)}\ket{r}\bra{r}
\label{POVM6Hmin}
	\\
	\ket{q} &=& -\sqrt{\frac{1-\qb}{3-4\qb}}\ket{m_0}+\frac{\sqrt{2-3\qb}}{\sqrt{3-4\qb}}
	\frac{\ket{m_1}+\ket{m_2}-2\ket{m_3}}{\sqrt6}
	\\
	\ket r &=& \sqrt{1-\qb}\frac{\ket{m_1}+\ket{m_2}+\ket{m_3} }{\sqrt3}
	+i\sqrt\qb\frac{\ket{m_1}-\ket{m_2}}{\sqrt2}
\label{defketr}
\eea
and $M_1=S M_3 S\dagg$, $M_2=SM_1 S\dagg$. 
\end{theorem}
\underline{Proof}:
For $b\in\{1,2,3\}$ we have
\bea
	\qz_b &=& (1-\fr32\qb)\ket{m_0}\bra{m_0}+\fr\qb2(\ket{m_1}\bra{m_1}+\ket{m_2}\bra{m_2}+\ket{m_3}\bra{m_3})
	\nn\\ &&
	+ \sqrt{\fr\qb2(1-\fr32\qb)}(\ket{m_0}\bra{m_b}+{\rm h.c.})
	+\fr\qb2(i\ket{m_{b+1}}\bra{m_{b+2}}+{\rm h.c.})
\eea
where $b+1$ should be read as $b+1\!\!\mod 3\in\{1,2,3\}$.

The matrix $\qL$ as defined in Section~\ref{sec:notation} is given by
\bea
	\qL &= &\sum_b \qz_b M_b = (1-\fr32\qb)(1+\frac{2\sqrt\qb}{\sqrt6\sqrt{1-\qb}})\ket{m_0}\bra{m_0}
	+(\frac12+\frac{(2-\qb)\sqrt\qb}{3\sqrt6\sqrt{1-\qb}})\sum_{j=1}^3 \ket{m_j}\bra{m_j}
	\\ &&
	+\frac{\sqrt2}{6}\sqrt{\qb(1-\qb)}\left[\sum_{j=1}^3\ket{m_0}\bra{m_j}+{\mbox h.c.}\right]
	+[(\frac{-i\qb}{2}-\frac{(1-2\qb)\sqrt\qb}{3\sqrt6\sqrt{1-\qb}})\sum_{j=1}^3\ket{m_{j+1}}\bra{m_j}
	+{\mbox h.c.}].
	\nn
\eea
With some effort it is verified that indeed $\qL-\qz_b\geq 0$ for $b\in\{1,2,3\}$ and $\qb\in[0,\fr12]$.
\hfill$\square$

\begin{conjecture}
\label{conj:K26Shannon}
Consider 6-state encoding.
In terms of Shannon entropy,
Eve's optimal POVM $\cQ=(Q_b)_{b\in\cB}$ for learning as much as possible about $B$ from $\qz_B$ is given by
\bea
	Q_3 &=&  \frac{3-4\qb}{3(1-\qb)}\ket{q'}\bra{q'}+\frac1{3(1-\qb)}\ket{r'}\bra{r'}
\label{POVM6Shannon}
	\\
	\ket{q'} &=& \sqrt{\frac{1-\qb}{3-4\qb}}\ket{m_0}+\frac{\sqrt{2-3\qb}}{\sqrt{3-4\qb}}
	\frac{\ket{m_1}+\ket{m_2}-2\ket{m_3}}{\sqrt6}
	\\
	\ket{r'} &=& \ket r^*
\eea
with $\ket r$ as defined by (\ref{defketr}),
and $Q_1=S Q_3 S\dagg$, $Q_2=S Q_1 S\dagg$.
\end{conjecture}
\underline{Evidence}:
The POVM $\cQ$ is the `dual' of $\cM$ in the sense that it has $\vecv$ replaced by $-\vecv$.
(This fact is not immediately evident. One can also take $\cM$ and apply it to the state $\qz_B$
with $\vecv\to-\vecv$; this is equivalent).
It was noticed in \cite{SdV2016} that such a `dual' is the optimal POVM in the case of 
the intercept attack~K1.
We have performed numerical POVM optimisations which find a local minimum of the Shannon entropy,
starting from $3^{10}$ initial points in POVM space;
all combinations of a positive/zero/negative value for each of the 10 degrees of freedom that
are left in the POVM after imposing $S$-symmetry.\footnote{
Imposing symmetry is allowed, see Section~\ref{sec:notation}.
}
Furthermore we did a Monte Carlo sampling of $10^{11}$ random POVMs.
We did not find a single POVM that performs better than~$\cQ$.
The numerical search did find $\cM$ and $\cQ$, as well as $200$ POVMs with Shannon entropy
between that of $\cQ$ and $\cM$.
\hfill$\square$

\begin{theorem}
In case of the measurement $\cQ$ specified in Conjecture~\ref{conj:K26Shannon},
the entropy of $B$ is given by 
\bea
	\sH(B|\cQ(\qz_B)) &=& h(p_6)+1-p_6
	\\
	p_6 &\isdef& \frac13-\frac{2\sqrt2}{3\sqrt3}\sqrt{\qb(1-\qb)}.
\label{K26Shannon}
\eea
\end{theorem}
\underline{Proof}:
After some algebra it can be seen that
$\tr \qz_3 Q_3=p_6$. We apply (\ref{K2hshort}) from Lemma~\ref{lemma:KShannon}.
\hfill$\square$

Some remarks on the case $\qb\geq\fr13$ can be found in the Appendix.

\subsection{Attack K2, 8-state}
\label{sec:K28}

\begin{theorem}
\label{th:K28Hmin}
Let $\qb\leq\fr13$.
In the 8-state case, the min-entropy of $B$ given the mixed state $\qz_B$ is
\be
	\Hmin(B|\qz_B)= \Hmin(B)-\log\left(1+\sqrt6\sqrt{\qb(1-\fr32\qb)}\right).
\ee
The associated POVM $(M_{uw})_{u,w\in\bits}$ is
\bea
	M_{00}=\frac{\sum_{a=0}^3\ket{m_a}}2\frac{\sum_{a'=0}^3\bra{m_{a'}}}2 &;&
	M_{01}= (\qs_z\otimes\one)M_{00}(\qs_z\otimes\one)
	\\
	M_{10}= (\qs_z\otimes\qs_z)M_{00}(\qs_z\otimes\qs_z)
	&;&
	M_{11}=(\one\otimes\qs_z)M_{00} (\one\otimes\qs_z).
\eea
\end{theorem}
\underline{Proof}:
The states $\qz_{uw}$ are given by
\bea
	\qz_{00} &=& (1-\fr32\qb)\ket{m_0}\bra{m_0}+\frac\qb2\sum_{j=1}^3 \ket{m_j}\bra{m_j}
	+\sqrt{\fr\qb2(1-\fr32\qb)}\left[\ket{m_0}\frac{\bra{m_1}+\bra{m_2}+\bra{m_3}}{\sqrt3} +{\rm h.c.}\right]
	\nn\\ &&
	+\frac\qb{2\sqrt3}\left[ i\sum_{j=1}^3\ket{m_{j}}\bra{m_{j+1}} + {\rm h.c.}
	\right]
\eea
and $\qz_{01}= (\qs_z\otimes\one)\qz_{00}(\qs_z\otimes\one)$,
$\qz_{10}= (\qs_z\otimes\qs_z)\qz_{00}(\qs_z\otimes\qs_z)$,
$\qz_{11}=(\one\otimes\qs_z)\qz_{00} (\one\otimes\qs_z)$.
The matrix $\qL$ has a simple diagonal form,
\be
	\qL = \sum_{uw}\qz_{uw}M_{uw}=
	\left(1-\fr32\qb+\sqrt3\sqrt{\fr\qb2(1-\fr32\qb)}\right)\ket{m_0}\bra{m_0}
	+(\frac\qb2+\frac{\sqrt{\fr\qb2(1-\fr32\qb)}}{\sqrt3})\sum_{j=1}^3\ket{m_j}\bra{m_j}.
\ee
It is easily verified that $\qL-\qz_{uw}\geq 0$ for all $\qb\in[0,\fr13]$ and $u,w\in\bits$.
Furthermore we have
\be
	\tr\qL=1+\sqrt6\sqrt{\qb(1-\fr32\qb)}.
\ee
\hfill$\square$

\begin{conjecture}
\label{conj:K28Shannon}
Consider 8-state encoding. Let $\qb\leq\fr13$.
In terms of Shannon entropy,
Eve's optimal POVM $\cR=(R_{uw})_{u,w\in\bits}$ for learning as much as possible about $U,W$ from $\qz_{UW}$ is given by
\be
	R_{00}=\ket v\bra v, \quad\quad
	\ket v=\frac{\ket{m_0}-\ket{m_1}-\ket{m_2}-\ket{m_3}}2 
\ee
and 
$R_{01}= (\qs_z\otimes\one)R_{00}(\qs_z\otimes\one),
	R_{10}= (\qs_z\otimes\qs_z)R_{00}(\qs_z\otimes\qs_z),
	R_{11}=(\one\otimes\qs_z)R_{00} (\one\otimes\qs_z)$.
\end{conjecture}
\underline{Evidence}:
Just as in the 6-state case,
the POVM for the Shannon entropy is the `dual' ($\vecv\to -\vecv$)
of the POVM associated with the min-entropy.
Numerical optimisations (from $3^{12}$ initial points) with imposed symmetry  
gave us no POVM that performs better than~$\cR$.
The numerical search did find $\cR$ and $\cM$, as well as 168 POVMs with Shannon entropy
between that of $\cR$ and $\cM$.
\hfill$\square$

\begin{theorem}
\label{th:K28Shannon}
In case of the measurement $\cR$ specified in Conjecture~\ref{conj:K28Shannon},
the entropy of $B$ is given by 
\bea
	\sH(B|\cR(\qz_B)) &=& h(p_8)+(1-p_8)\log3
\label{K28Shannon}
	\\
	p_8 &\isdef& \frac14-\frac{\sqrt3}{2\sqrt2}\sqrt{\qb(1-\fr32\qb)}.
\label{defp8}
\eea
\end{theorem}
\underline{Proof}:
A brief calculation gives $\tr \qz_{uw}R_{uw}=p_8$ (for all $u,w$)
with $p_8$ as defined in (\ref{defp8}). Then we use (\ref{K2hshort}).
\hfill$\square$

Some remarks on the case $\qb\geq\fr13$ can be found in the Appendix.

\section{Putting it all together}
\label{sec:together}

The amount of privacy amplification needed in the protocol
(Section~\ref{sec:QKR}, {\tt Ext} function)
is determined by the {\em strongest} of the M1, M2, K1, K2 attacks.
Below we combine all the results from Sections \ref{sec:message} and \ref{sec:key}.

\begin{table}[h]
\begin{tabular}{|c|c|c|c|}
\hline
\multicolumn{4}{|c|}{\bf Shannon entropy leakage $I(\qb)$ per qubit}
\\ \hline
&{\it 4-state}& {\it 6-state}& {\it 8-state}
\\ \hline
M1 & 0.399 & 0.256 & 0
\\ \hline
M2& \multicolumn{3}{|c|}{$\qb\cdot1+(1-\qb)[1-h(p_\qb)]$,\;\; $p_\qb=\frac12-\frac{\sqrt{\fr\qb2(1-\fr32\qb)}}{1-\qb}$}
\\ \hline
K1& $3\qb\cdot 0.399$ & $3\qb\cdot 0.314$ & $3\qb\cdot 0.415$
\\ \hline
K2& $1-h(\frac12+\frac12\sqrt{\qb(1-\fr32\qb)}+\frac{\qb}{2\sqrt2})$ & 
$\log3-[h(p_6)+1-p_6]$ 
 & $2-[h(p_8)+(1-p_8)\log3]$
\\ 
&& $p_6=\frac13-\frac{2\sqrt2}{3\sqrt3}\sqrt{\qb(1-\qb)}$ &
$p_8=\fr14-\fr{\sqrt6}{4}\sqrt{\qb(1-\fr32\qb)}$
\\ \hline
\end{tabular}
\caption{\it Shannon entropy loss $I(\qb)$ as a function of noise $\qb$, for the attacks M1,M2,K1,K2.
The 6-state and 8-state K2 results are conjectures.}
\label{t:Shannonloss}
\end{table}

\begin{figure}[h]
\begin{center}
\setlength{\unitlength}{1mm}
\begin{picture}(140,32)(0,0)
\put(0,0){\includegraphics[width=45mm]{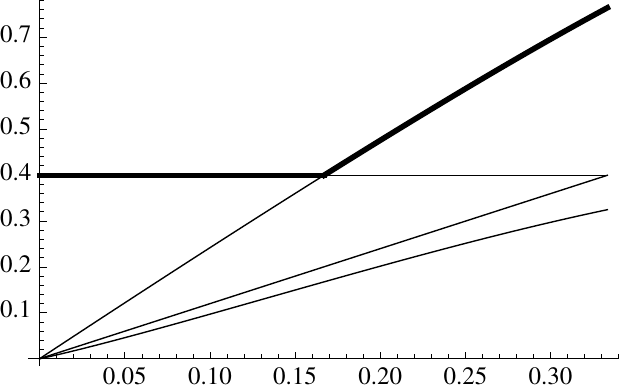}}
\put(48,0){\includegraphics[width=45mm]{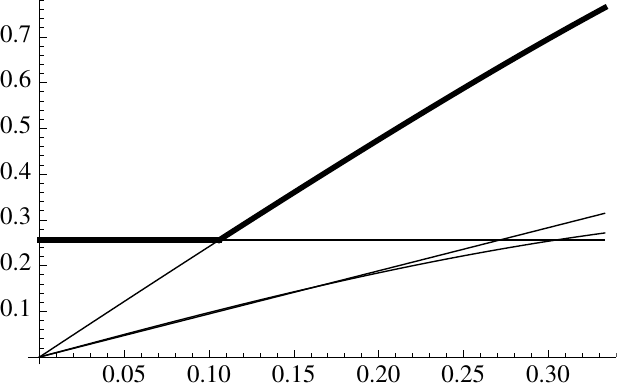}}
\put(96,0){\includegraphics[width=45mm]{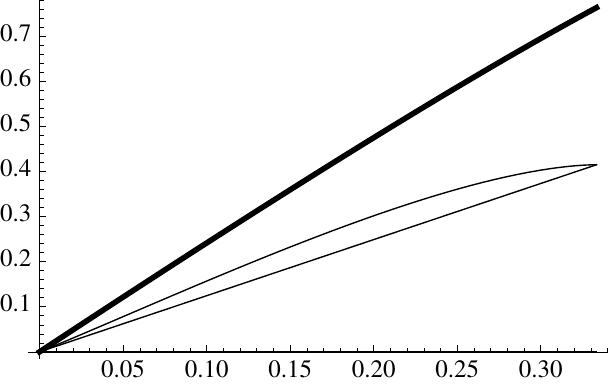}}
\put(43,-1){$\qb$}
\put(91,-1){$\qb$}
\put(139,-1){$\qb$}
\put(2,30){\small{$I$}}
\put(50,30){\small{$I$}}
\put(98,30){\small{$I$}}
\put(18,30){\fbox{\small{4-state}}}
\put(66,30){\fbox{\small{6-state}}}
\put(114,30){\fbox{\small{8-state}}}
\put(5,17){\tiny{M1}}
\put(53,12){\tiny{M1}}
\put(20,9){\tiny{K1}}
\put(89,14){\tiny{K1}}
\put(116,6){\tiny{K1}}
\put(41,29){\tiny{M2}}
\put(89,29){\tiny{M2}}
\put(127,23){\tiny{M2}}
\put(29,8){\tiny{K2}}
\put(83,8){\tiny{K2}}
\put(127,15){\tiny{K2}}
\end{picture}
\caption{\it Shannon leakage $I(\qb)$ per qubit as a function of the bit error rate $\qb$.
The 6-state and 8-state K2 results are conjectures.}
\label{fig:leaksShannon}
\end{center}
\end{figure}
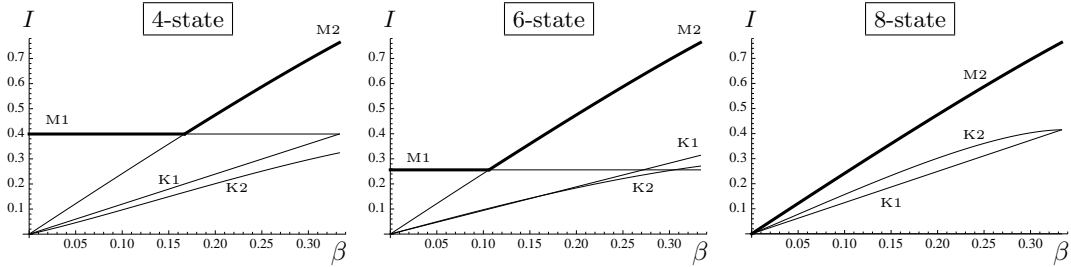

\begin{figure}[!h]
\begin{center}
\setlength{\unitlength}{1mm}
\begin{picture}(90,55)(0,0)
\put(0,0){\includegraphics[width=90mm]{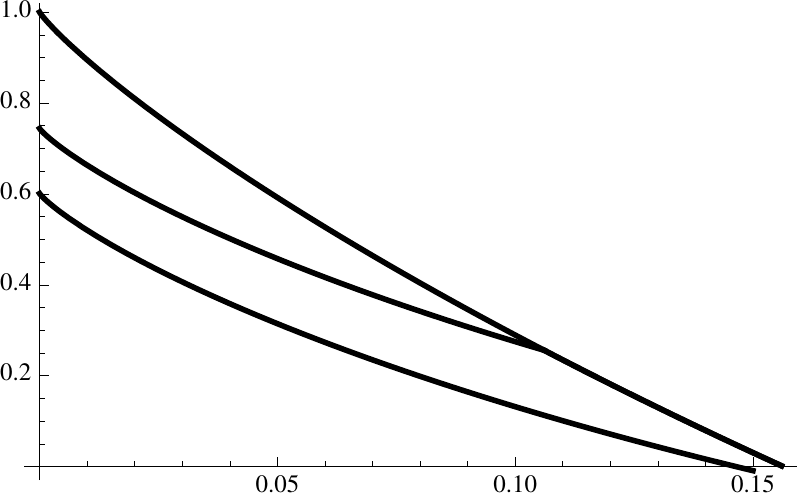}}
\put(1,2){\small{$0$}}
\put(89,0){$\qb$}
\put(8,54){$1-h(\qb)-\max I(\qb)$}
\put(42,9){\small{4-state}}
\put(10,38){\small{6-state}}
\put(34,34){\small{8-state}}
\end{picture}
\caption{\it QKR capacity $1-h(\qb)-\max_{\rm attacks} I(\qb)$
as a function of the bit error rate $\qb$. (Leakage is expressed as mutual information).
The strongest attack determines $I(\qb)$.}
\label{fig:capacityShannon}
\end{center}
\end{figure}
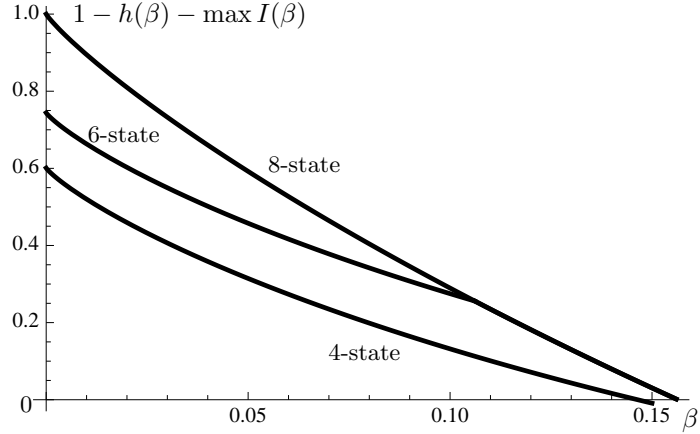

\subsection{Combined results for Shannon entropy}
\label{sec:comboShannon}

Table~\ref{t:Shannonloss} shows an overview of the Shannon entropy losses 
in all the attacks.
The individual M1,M2,K1,K2 leakages (and the maximum) are plotted as a function of $\qb$
in Fig.\;\ref{fig:leaksShannon}. 
Fig.\;\ref{fig:capacityShannon} shows the QKR capacity $1-h(\qb)-I(\qb)$.

For 4-state and 6-state encoding, the strongest attack at low $\qb$ is~M1.
At larger $\qb$ it is the QKD-like attack M2.
For 8-state encoding, M2 is always the strongest attack.
The QKR channel capacity of 4-state encoding is always below 6-state.
8-state has higher capacity than 6-state
at $\qb$ up to $\approx 0.1061$, after which they are the same and equal to the QKD capacity.

Our plots do not go beyond $\qb=\fr13$ because intercept-resend attacks cause noise $\qb=\fr13$.
In attack K1 the fraction of qubits intercepted by Eve is $3\qb$, which at $\qb>\fr13$ would exceed~$1$.
At $\qb>\fr13$ we have to be careful how we interpret K1. 
A discussion can be found in the Appendix.
Note that attacks K1 and K2 at $\qb=\fr13$ are not necessarily the same thing.
Attack K2 restricts Eve's options by forcing her to first perform a specific ancilla operation,
whereas attack K1 allows any POVM on the intercepted qubit.
Hence at $\qb=\fr13$ the K2 leakage cannot exceed the K1 leakage.

\subsection{Combined results for min-entropy}
\label{sec:comboHmin}

\begin{figure}[h]
\begin{center}
\setlength{\unitlength}{1mm}
\begin{picture}(140,32)(0,0)
\put(0,0){\includegraphics[width=45mm]{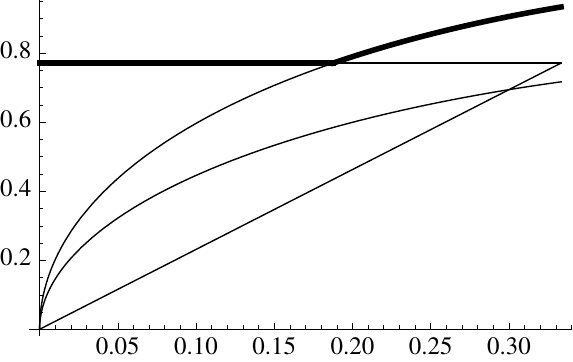}}
\put(48,0){\includegraphics[width=45mm]{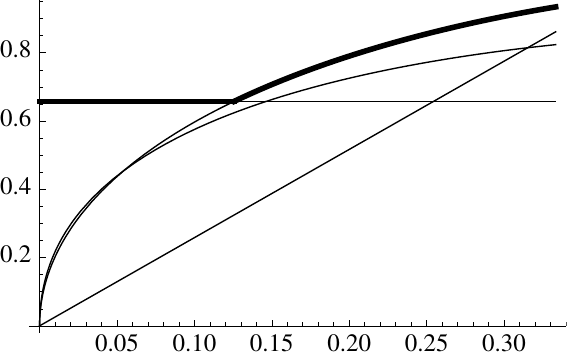}}
\put(96,0){\includegraphics[width=45mm]{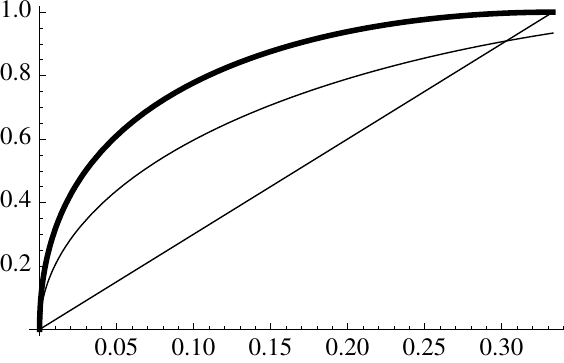}}
\put(43,-1){$\qb$}
\put(91,-1){$\qb$}
\put(139,-1){$\qb$}
\put(0,30){\small{$\tri\Hmin$}}
\put(48,30){\small{$\tri\Hmin$}}
\put(96,30){\small{$\tri\Hmin$}}
\put(18,30){\fbox{\small{4-state}}}
\put(66,30){\fbox{\small{6-state}}}
\put(114,30){\fbox{\small{8-state}}}
\put(5,21){\tiny{M1}}
\put(53,18){\tiny{M1}}
\put(20,9){\tiny{K1}}
\put(67,9){\tiny{K1}}
\put(114,9){\tiny{K1}}
\put(41,29){\tiny{M2}}
\put(89,29){\tiny{M2}}
\put(108,14){\tiny{M2}}
\put(17,14){\tiny{K2}}
\put(66,17){\tiny{K2}}
\put(105,21){\tiny{K2}}
\end{picture}
\caption{\it Min-entropy leakage per qubit as a function of the bit error rate $\qb$.}
\label{fig:leaksHmin}
\end{center}
\end{figure}
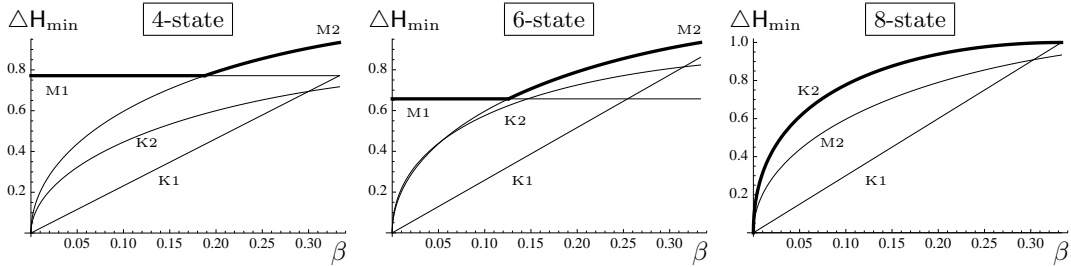

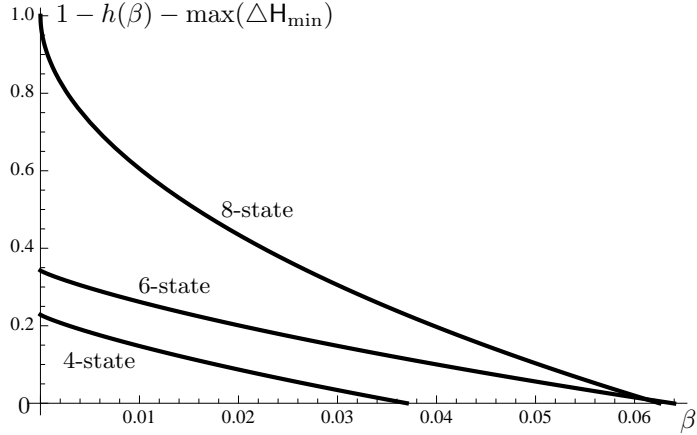
\begin{figure}[h]
\begin{center}
\setlength{\unitlength}{1mm}
\begin{picture}(90,55)(0,0)
\put(0,0){\includegraphics[width=90mm]{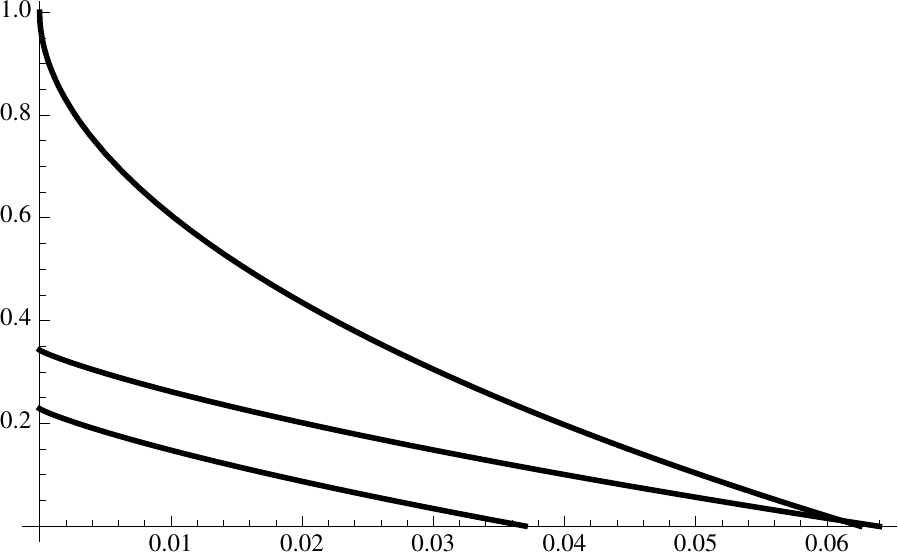}}
\put(1,2){\small{$0$}}
\put(89,0){$\qb$}
\put(6,54){$1-h(\qb)-\max(\tri\Hmin)$}
\put(7,8){\small{4-state}}
\put(17,18){\small{6-state}}
\put(28,28){\small{8-state}}
\end{picture}
\caption{\it QKR capacity as a function of the bit error rate $\qb$, if leakage is expressed as min-entropy loss.}
\label{fig:capacityHmin}
\end{center}
\end{figure}

Table~\ref{t:Hminloss} shows an overview of the min-entropy entropy losses 
in all the attacks.
The individual M1,M2,K1,K2 leakages (and the maximum) are plotted as a function of $\qb$
in Fig.\;\ref{fig:leaksHmin}. 
Fig.\;\ref{fig:capacityHmin} shows the QKR capacity $1-h(\qb)-\tri\Hmin(\qb)$.
For 4-state and 6-state, the winning attacks are as for the Shannon entropy case.
For 8-state, however, the winning attack is~K2.
If capacity is computed using min-entropy loss as the measure of Eve's knowledge,
then the QKR capacity of 8-state is higher than 6-state on the range
$\qb\in[0,0.0612]$.
There is a tiny interval $\qb\in(0.0612, 0.0638]$ where 6-state outperforms 8-state;
at $\qb>0.0638$ all capacities are zero.
4-state is always worse than 6-state.

\begin{table}[h]
\begin{tabular}{|c|c|c|c|}
\hline
\multicolumn{4}{|c|}{\bf Min-entropy leakage per qubit}
\\ \hline
&4-state& 6-state& 8-state
\\ \hline
M1 & 0.772 & 0.658 & 0
\\ \hline
M2& \multicolumn{3}{|c|}{$\log[1+\sqrt2\sqrt{\qb(1-\fr32\qb)}+\qb]$}
\\ \hline
K1& $3\qb\cdot 0.772$ & $3\qb\cdot 0.861$ & $3\qb\cdot 1$
\\ \hline
K2& $\log(1+\sqrt{\qb(1-\fr32\qb)}+\frac\qb{\sqrt2})$ & $\log\left(1+\frac{2\sqrt2}{\sqrt3}\sqrt{\qb(1-\qb)}\right)$ &
$\log\left(1+\sqrt6\sqrt{\qb(1-\fr32\qb)}\right)$
\\ \hline
\end{tabular}
\caption{\it Min-entropy loss as a function of noise $\qb$, for the attacks M1,M2,K1,K2.}
\label{t:Hminloss}
\end{table}


\section{Addition of artificial noise}
\label{sec:QKDwithnoise}

The structure evident in the $\ket{E^\vecv_{xy}}$ vectors (\ref{vvbasis})
allows us to simplify the derivation of the capacity of 6-state/8-state QKD with added
artificial noise.
(This also applies to attack M2.)
In \cite{SKMB2008} a derivation for 6-state QKD was given without noise symmetrisation, resulting 
in a lengthy analysis. Moreover, the end result was presented in a less than elegant way.
Here we give a shorter derivation, and we present the end result in a very intuitive form.

\begin{figure}[h]
\begin{center}
\setlength{\unitlength}{1mm}
\begin{picture}(140,40)(0,0)
\put(0,0){\includegraphics[width=65mm]{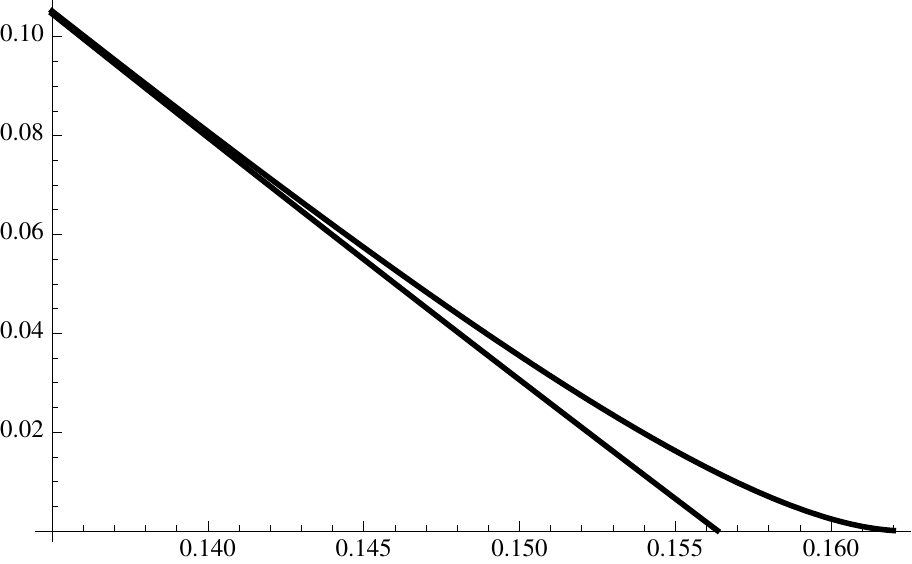}}
\put(70,0){\includegraphics[width=65mm]{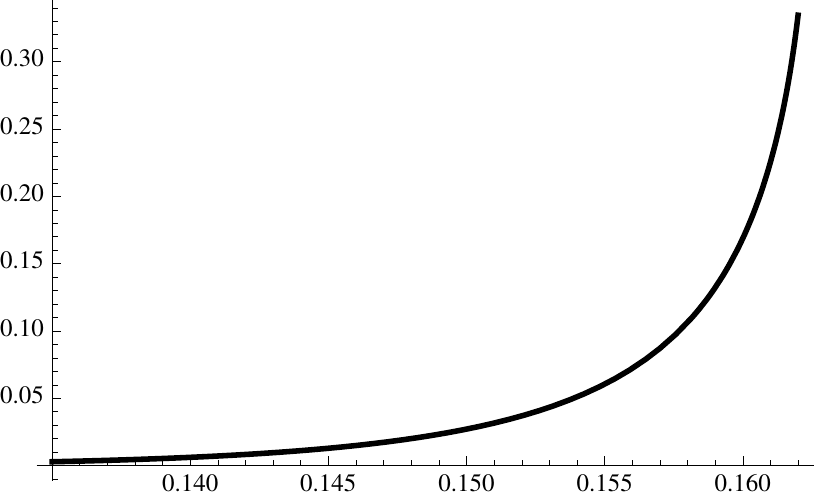}}
\put(65,0){\small $\qb$}
\put(135,0){\small $\qb$}
\put(42,4){$C$}
\put(55,6){$C_{\rm opt}$}
\put(75,38){$\qe_{\rm opt}$}
\end{picture}
\caption{\it {\bf Left:} The capacity $C(\qb)$ without artificial noise and 
the capacity $C_{\rm opt}=C'(\qe_{\rm opt}(\qb),\qb)$ for the best choice of artificial noise.
{\bf Right:} The optimal value of Alice's noise parameter $\qe$ as a function of the channel noise~$\qb$.
(Numerical optimisation.)}
\label{fig:epsilon}
\end{center}
\end{figure}
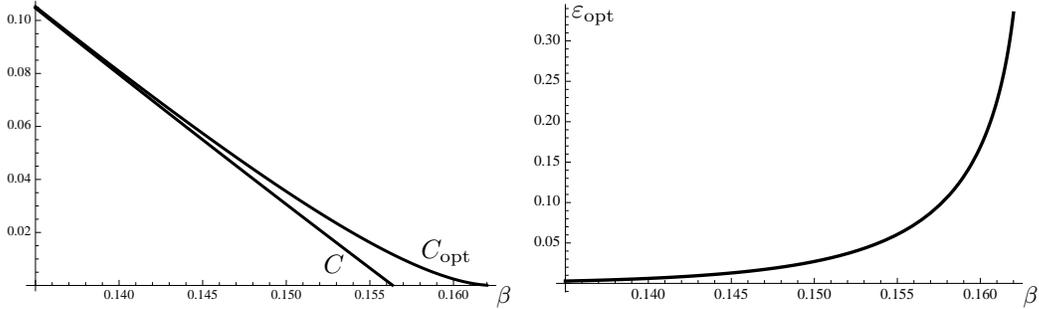

Alice adds artificial noise to $X$. This is represented as a binary symmetric channel with
bit error rate~$\qe$.
Let $\qe\star\qb\isdef\qe(1-\qb)+(1-\qe)\qb$ be the bit error rate on the concatenated
channel consisting of Alice's noise $\qe$ followed by the physical noise $\qb$ introduced by Eve.
The channel capacity from Alice to Bob becomes $I_{\rm AB}'(\qe,\qb)=1-h(\qe\star\qb)$.
Eve's task of distinguishing between the various $\ket{E^\vecv}$ states
is not affected; the weights $\qb$ and $1-\qb$ 
in (\ref{IAEQKD})
do not change.
However, Eve's inference about $X$ from her measurement outcomes has additional noise~$\qe$:
the bit error rate of the `easy' channel changes from $0$ to $\qe\star 0=\qe$,
and the bit error rate of the `difficult' channel changes from $p_\qb$ to $\qe\star p_\qb$.
Thus the channel from Alice to Eve now has capacity
$I_{\rm AE}'(\qe,\qb)=\qb[1-h(\qe)]+(1-\qb)[1-h(\qe\star p_\qb)]$,
with $p_\qb$ as defined in (\ref{defpbeta}).
The secrecy capacity is
\bea
	C'(\qe,\qb)=I_{AB}'-I_{AE}' &=& 1-h(\qe\star\qb)-\left\{
	\qb[1-h(\qe)]+(1-\qb)[1-h(\qe\star p_\qb)]
	\vphantom{M^{M^M}}\right\}
	\nn\\ &=&
	(1-\qb)h(\qe\star p_\qb)+\qb h(\qe)-h(\qe\star\qb)
\eea
which is precisely the result of \cite{SKMB2008} but in simplified form.
Fig.\;\ref{fig:epsilon} shows the optimal noise $\qe_{\rm opt}(\qb)$ as a function of $\qb$, and the resulting capacity
$C_{\rm opt}(\qb)=C'(\qe_{\rm opt}(\qb),\qb)$.
The original positive-capacity region $\qb\leq 0.156$
is extended to $\qb\leq0.162$.

\section{Discussion}
\label{sec:discussion}

The fact that M1 is the dominant attack against 4-state and 6-state encoding
at low bit error rate, and M2 at larger $\qb$, comes as no surprise.
The vulnerability of the message 
is exactly the reason why 8-state encoding was introduced in \cite{SdV2016}.
And as 8-state protects the message better,
it is also not surprising that an attack on the key dominates in the 8-state min-entropy analysis.

What we did not know a priori is the relative strength of the $\qb$-dependent attacks,
and their strength (at large $\qb$) compared to~M1. 
Figs.\;\ref{fig:leaksShannon} and \ref{fig:leaksHmin} show complicated behaviour with 
various intersections of curves.

We were surprised to see M2 `winning' in the 8-state Shannon entropy analysis.
With M2 being the relevant attack, a large part of the security analysis
becomes identical, or at least very similar, to well known QKD analysis.
Hence the trick with Alice's artificial noise is as relevant to QKR as it is to QKD.

When the number of qubits ($n$) is very large, the relevant quantity to look at is
Shannon entropy. For small $n$ it is min-entropy. In intermediate cases it is something in between.
From our results we conclude that 8-state encoding yields the highest QKR capacity
under practically all circumstances. 

\vskip1mm

As topics for future work we see
(i)
Adaptation of the protocol so that the $n$-qubit
quantum state $\ket\qJ$ sent by Alice contains the message itself
(in privacy-amplified form, as in \cite{uncl}), instead of a random mask.
This would further improve communication efficiency.
(ii)
Determine the effect of artificial noise on the min-entropy loss in the case of the K2 attack
on 8-state encoding.
(iii)
Determine how tight the bound in Lemma~\ref{lemma:equivQKD} (M2 reduces to QKD analysis)
is as a function of~$N$.

\section*{Acknowledgments}

We thank Serge Fehr for useful discussions.
Part of this research was funded by NWO (CHIST-ERA project ID\underbar{\;\;}IOT). 

\appendix

\section*{Appendix: Attack K2 at high noise levels}

For the sake of completeness we present entropy results for the K2 attack at very high noise levels. 
As mentioned in Section~\ref{sec:comboShannon}, the K1 attack needs some interpreting at $\qb>\fr13$:
Eve does the the optimal K1-POVM on all $n$ qubits but then forwards badly chosen
states to Bob which cause $\qb>\fr13$.
Attack K2 is still defined as before: Eve couples her ancilla to the AB system in such a way that
noise $\qb>\fr13$ occurs. 
At $\qb=\fr12$ the point is reached where Eve might as well send a completely random qubit state to Bob,
and she extracts the maximum possible amount of information from the scrutinised qubit.
Hence the K2-leakage at $\qb=\fr12$ must equal the K1-leakage at $\qb=\fr13$. 

In the case of 4- and 6-state encoding we find that the POVMs (\ref{POVM4Hmin}) 
and (\ref{POVM6Hmin},\ref{POVM6Shannon}) respectively
are optimal on the whole range $\qb\in[0,\fr12]$.
In the 8-state case the situation is different:
we find a different POVM in the range $\qb\in[\fr13,\fr12]$.

\begin{theorem}
\label{th:K28Hmin_noisy}
Let $\frac{1}{3}\leq\qb\leq\fr12$.
For 8-state encoding, the min-entropy of $B$ given the mixed state $\qz_B$ is
\be
	\Hmin(B|\qz_B)= \Hmin(B) -1=1.
\ee
The associated POVM $(M_{uw})_{u,w\in\bits}$ is
\bea
	M_{00} &=& \frac{1-\qb}{2\qb}\ket{a}\bra{a}+\frac{3\qb-1}{2\qb}\ket{d}\bra{d}
	\\
	\ket{a} &=& \frac{\sqrt{\qb/2}}{\sqrt{1-\qb}} \ket{m_0}
	+\frac{\sqrt{1-\fr32\qb}}{\sqrt{1-\qb}}\cdot\frac{\ket{m_1}+\ket{m_2}+\ket{m_3}}{\sqrt3}
	\\
	\ket{d} &=& \frac{e^{i\pi/3} \ket{m_1} + e^{-i\pi/3} \ket{m_2} - \ket{m_3}}{\sqrt3}
\eea
\bea
	M_{01}= (\qs_z\otimes\one)M_{00}(\qs_z\otimes\one) ;
	M_{10}= (\qs_z\otimes\qs_z)M_{00}(\qs_z\otimes\qs_z) ;
	M_{11}=(\one\otimes\qs_z)M_{00} (\one\otimes\qs_z).
\eea

\end{theorem}
\underline{Proof}:
After some algebra it turns out that
the matrix $\qL$ has a simple diagonal form,
\be
	\qL = \sum_{uw}\qz_{uw}M_{uw}=
	\left(2-3\qb\right)\ket{m_0}\bra{m_0}
	+\qb\sum_{j=1}^3\ket{m_j}\bra{m_j}.
\ee
It is easily verified that $\qL-\qz_{uw}\geq 0$ for all $\qb\in[\fr13,\fr12]$ and $u,w\in\bits$.
\hfill$\square$

\begin{lemma}
\label{lemma:K28Shannon_noisy}
Consider 8-state encoding. Let $\fr13 \leq \qb \leq \fr12$.
In terms of Shannon entropy,
Eve's optimal POVM $\cR=(R_{uw})_{u,w\in\bits}$ for learning as much as possible about $U,W$ from $\qz_{UW}$ is given by
\bea
	R_{00} &=& \frac{1-\qb}{2\qb}\ket{a'}\bra{a'}+\frac{3\qb-1}{2\qb}\ket{d'}\bra{d'}
	\\
	\ket{a'} &=& -\frac{\sqrt{\qb/2}}{\sqrt{1-\qb}} \ket{m_0}
	+\frac{\sqrt{1-\fr32\qb}}{\sqrt{1-\qb}}\cdot\frac{\ket{m_1}+\ket{m_2}+\ket{m_3}}{\sqrt3}
	\\
	\ket{d'} &=& \ket{d}^{*}
\eea
and 
$R_{01}= (\qs_z\otimes\one)R_{00}(\qs_z\otimes\one),
	R_{10}= (\qs_z\otimes\qs_z)R_{00}(\qs_z\otimes\qs_z),
	R_{11}=(\one\otimes\qs_z)R_{00} (\one\otimes\qs_z)$.
\end{lemma}
\underline{Proof}:
On the whole range $\qb\in[\fr13,\fr12]$ the POVM $\cR$
gives $\sH(B|\cR(\qz_B)) = \log3$, which is the K1 result at $\qb=\fr13$
and therefore the minimum possible value.
\hfill$\square$

Just as in the 6-state case and in the 8-state for $\qb \leq \frac{1}{3}$,
the POVM $\cR$ for the Shannon entropy is the `dual' ($\vecv\to -\vecv$)
of the POVM associated with the min-entropy.

\begin{figure}[h]
\begin{center}
\setlength{\unitlength}{1mm}
\begin{picture}(140,32)(0,0)
\put(0,0){\includegraphics[width=45mm]{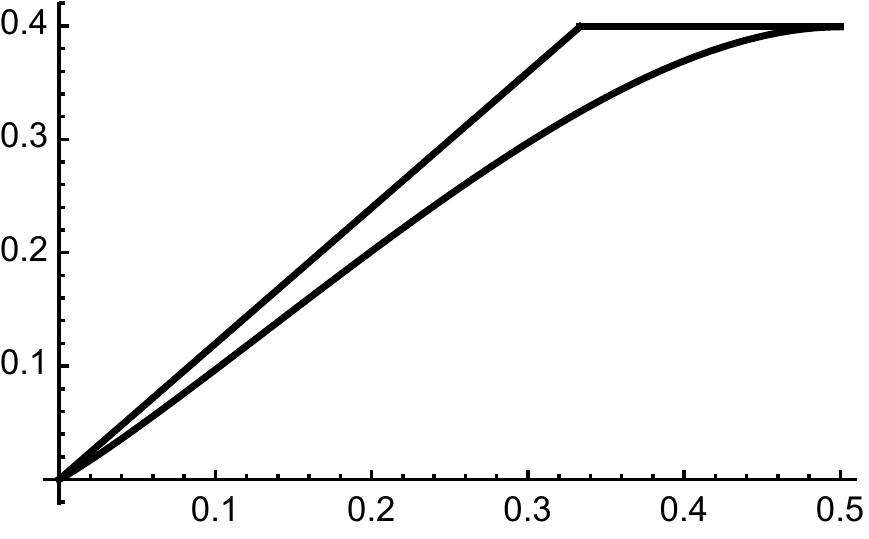}}
\put(48,0){\includegraphics[width=45mm]{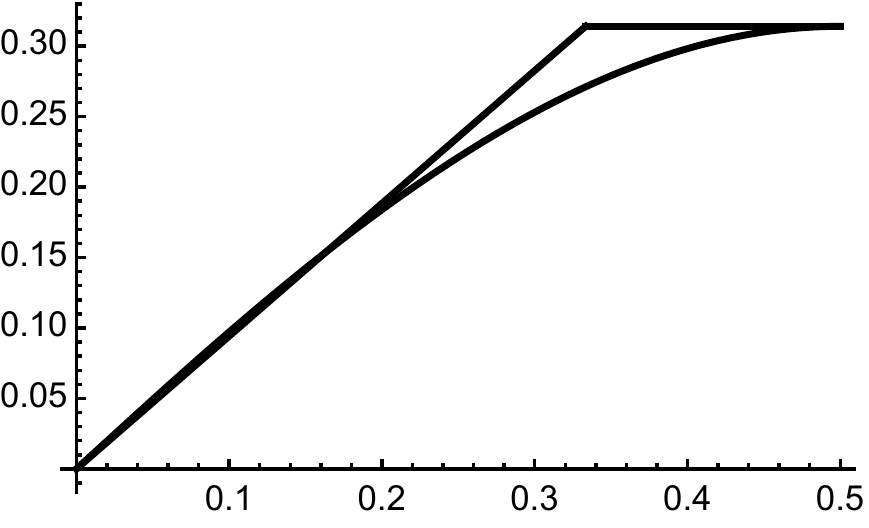}}
\put(96,0){\includegraphics[width=45mm]{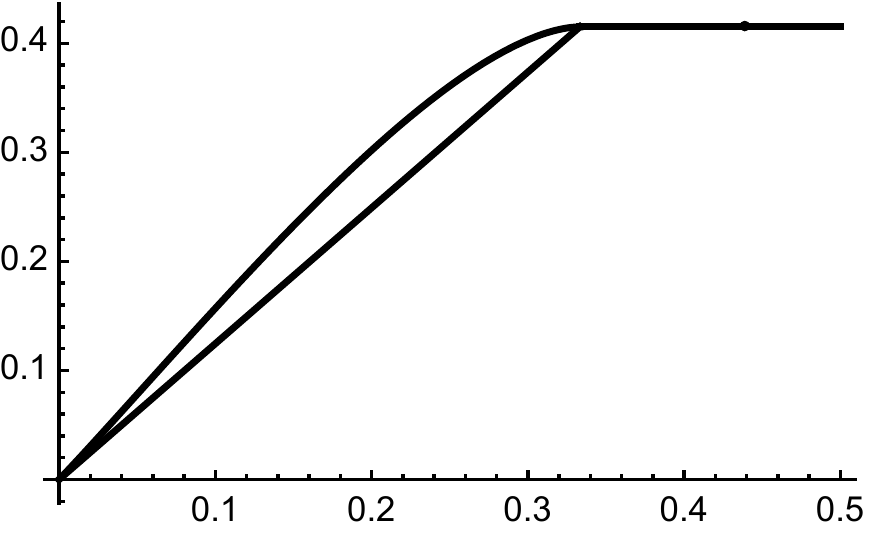}}
\put(43,-2){\small $\qb$}
\put(91,-2){\small $\qb$}
\put(139,-2){\small $\qb$}
\put(2,30){\small{$I$}}
\put(50,30){\small{$I$}}
\put(98,30){\small{$I$}}
\put(18,30){\fbox{\small{4-state}}}
\put(66,30){\fbox{\small{6-state}}}
\put(114,30){\fbox{\small{8-state}}}
\put(19,21){\tiny{K1}}
\put(72,24){\tiny{K1}}
\put(119,18){\tiny{K1}}
\put(30,20){\tiny{K2}}
\put(80,21){\tiny{K2}}
\put(112,21){\tiny{K2}}
\end{picture}
\caption{\it Shannon leakage $I(\qb)$ per qubit as a function of the bit error rate $\qb$ up to $\qb=\frac{1}{2}$.
The K2 results for 6-state and 8-state encoding are conjectures.}
\label{fig:noisyShannon}
\end{center}
\end{figure}
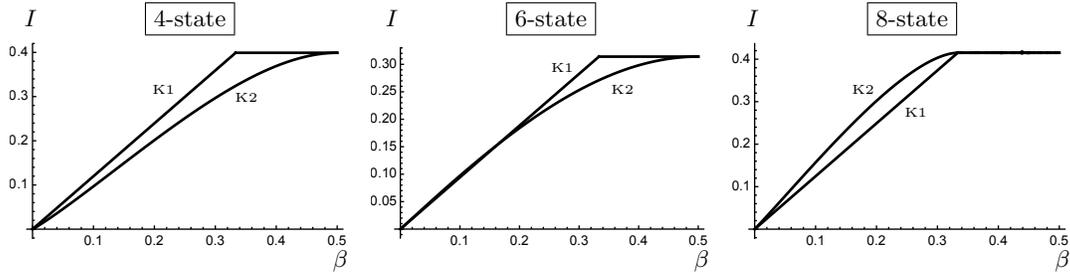

\begin{figure}[h]
\begin{center}
\setlength{\unitlength}{1mm}
\begin{picture}(140,32)(0,0)
\put(0,0){\includegraphics[width=45mm]{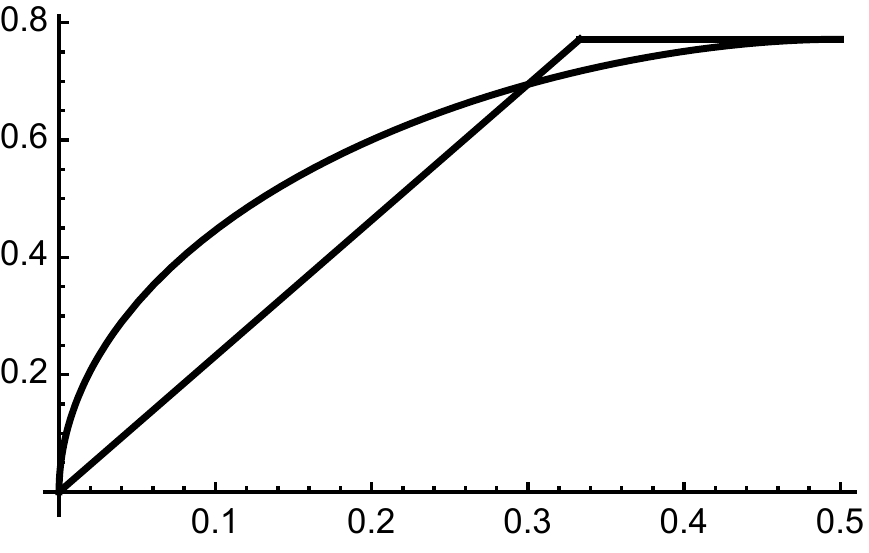}}
\put(48,0){\includegraphics[width=45mm]{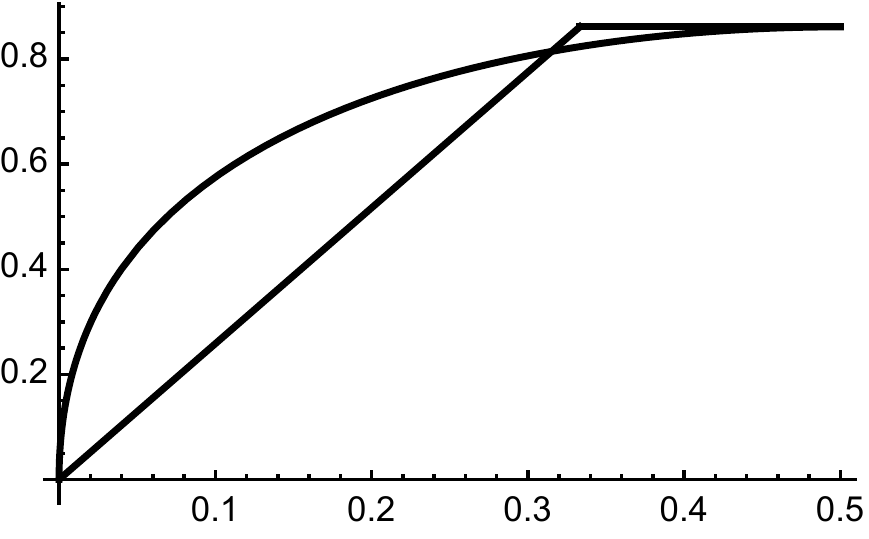}}
\put(96,0){\includegraphics[width=45mm]{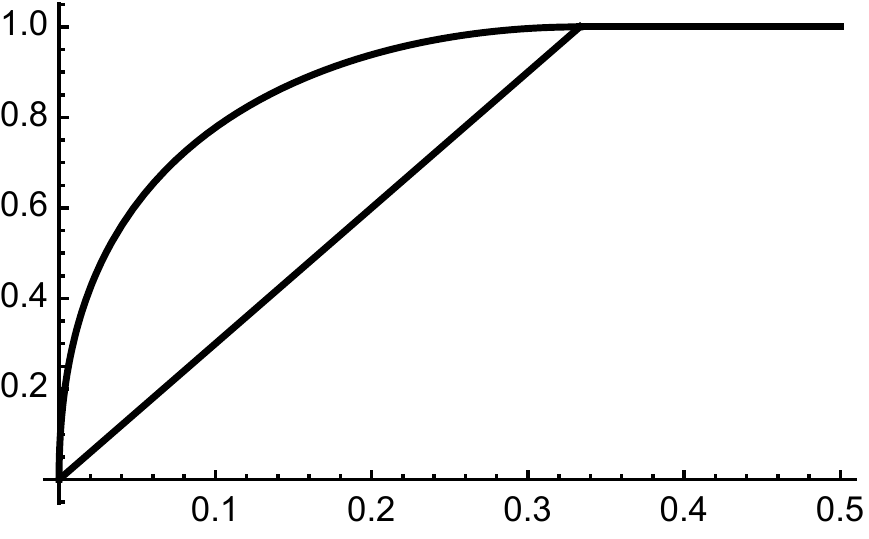}}
\put(43,-2){\small $\qb$}
\put(91,-2){\small $\qb$}
\put(139,-2){\small $\qb$}
\put(0,30){\small{$\tri\Hmin$}}
\put(48,30){\small{$\tri\Hmin$}}
\put(96,30){\small{$\tri\Hmin$}}
\put(18,30){\fbox{\small{4-state}}}
\put(66,30){\fbox{\small{6-state}}}
\put(114,30){\fbox{\small{8-state}}}
\put(17,12){\tiny{K1}}
\put(65,12){\tiny{K1}}
\put(113,12){\tiny{K1}}
\put(8,18){\tiny{K2}}
\put(53,18){\tiny{K2}}
\put(102,22){\tiny{K2}}
\end{picture}
\caption{\it Min-entropy leakage per qubit as a function of the bit error rate $\qb$ up to $\qb=\frac{1}{2}$.}
\label{fig:noisyHmin}
\end{center}
\end{figure}
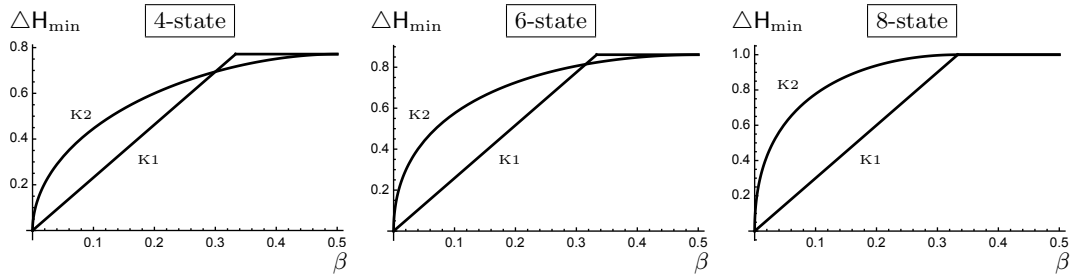

Note that at $\qb=\fr13$ the POVMs for $\qb\leq\fr13$ and $\qb\geq\fr13$ match, as they should.
The leakages for the K1 and K2 attacks up to $\qb = \frac{1}{2}$
are plotted in Figs. \ref{fig:noisyShannon} and \ref{fig:noisyHmin}.

For 4- and 6-state, K2 reaches it maximum at $\qb=\fr12$, whereas
in the 8-state case the maximum is reached already at $\qb=\fr13$.


\bibliographystyle{unsrt}

\bibliography{recyclingproof}

\end{document}